\documentclass[apj,numberedappendix]{emulateapj}
\usepackage{amssymb,amsmath}
\usepackage{graphicx}
\usepackage{appendix}
\usepackage{natbib}
\usepackage[backref,breaklinks,colorlinks,citecolor=blue]{hyperref}
\usepackage{enumerate}
\usepackage{multirow}
\usepackage{bm}

\altaffiltext{\MIT}{Kavli Institute for Astrophysics and Space Research, Massachusetts Institute of Technology, 77 Massachusetts Avenue, Cambridge, MA 02139}
\altaffiltext{\KIPAC}{Kavli Institute for Particle Astrophysics and Cosmology, Stanford University, 452 Lomita Mall, Stanford, CA 94305}
\altaffiltext{\Stanford}{Department of Physics, Stanford University, 382 Via Pueblo Mall, Stanford, CA 94305}
\altaffiltext{\SLAC}{SLAC National Accelerator Laboratory, 2575 Sand Hill Road, Menlo Park, CA 94025}
\altaffiltext{\UMontreal}{D\'{e}partement de Physique, Universit\'{e} de Montr\'{e}al, C.P. 6128, Succ. Centre-Ville, Montr\'{e}al, Qu\'{e}bec H3C 3J7, Canada}
\altaffiltext{\FNAL}{Fermi National Accelerator Laboratory, Batavia, IL 60510-0500, USA}
\altaffiltext{\KICPChicago}{Kavli Institute for Cosmological Physics, University of Chicago, Chicago, IL, USA 60637}
\altaffiltext{\AAUChicago}{Department of Astronomy and Astrophysics, University of Chicago, Chicago, IL, USA 60637}
\altaffiltext{\Miss}{Department of Physics and Astronomy, University of Missouri, 5110 Rockhill Road, Kansas City, MO 64110}
\altaffiltext{\CfA}{Harvard-Smithsonian Center for Astrophysics, 60 Garden Street, Cambridge, MA 02138}
\altaffiltext{\ASIAA}{Academia Sinica Institute of Astronomy and Astrophysics, 11F of AS/NTU Astronomy-Mathematics Building, No.1, Sec. 4, Roosevelt Rd, Taipei 10617, Taiwan}
\altaffiltext{\LMU}{Faculty of Physics, Ludwig-Maximilians-Universit\"{a}t, Scheinerstr.\ 1, 81679 Munich, Germany}
\altaffiltext{\ECUniverse}{Excellence Cluster Universe, Boltzmannstr.\ 2, 85748 Garching, Germany}
\altaffiltext{\Huntingdon}{Huntingdon Institute for X-ray Astronomy, LLC}
\altaffiltext{\Melbourne}{School of Physics, University of Melbourne, Parkville, VIC 3010, Australia}
\altaffiltext{\MPE}{Max Planck Institute for Extraterrestrial Physics, Giessenbachstr. 1, 85748 Garching, Germany}
\altaffiltext{\Bonn}{Argelander-Institut fur Astronomie, Auf dem Hugel 71, D-53121 Bonn, Germany}

\def\MIT{1}
\def\KIPAC{2}
\def\Stanford{3}
\def\SLAC{4}
\def\UMontreal{5}
\def\FNAL{6}
\def\KICPChicago{7}
\def\AAUChicago{8}
\def\Miss{9}
\def\CfA{10}
\def\ASIAA{11}
\def\LMU{12}
\def\ECUniverse{13}
\def\Huntingdon{14}
\def\Melbourne{15}
\def\MPE{16}
\def\Bonn{17}

\begin{document}


\title{A Detailed Study of the Most Relaxed SPT-Selected Galaxy Clusters: \\Cool Core and Central Galaxy Properties}
   
\author{
M.\,McDonald\altaffilmark{\MIT},
S.\,W.\,Allen\altaffilmark{\KIPAC,\Stanford,\SLAC},
J.\,Hlavacek-Larrondo\altaffilmark{\UMontreal},
A.\,B.\,Mantz\altaffilmark{\KIPAC,\Stanford,\SLAC},
M.\,Bayliss\altaffilmark{\MIT},
B.\,A.\,Benson\altaffilmark{\FNAL,\KICPChicago,\AAUChicago},
M.\,Brodwin\altaffilmark{\Miss},
E.\,Bulbul\altaffilmark{\CfA},
R.\ E.\ A.\ Canning\altaffilmark{\KIPAC,\Stanford},
I.\ Chiu \altaffilmark{\ASIAA,\LMU,\ECUniverse},
W.\ R.\ Forman\altaffilmark{\CfA},
G.~P.~Garmire\altaffilmark{\Huntingdon},
N.~Gupta\altaffilmark{\Melbourne},
G.~Khullar\altaffilmark{\AAUChicago},
J.~J.~Mohr\altaffilmark{\LMU,\ECUniverse,\MPE},
C.~L.~Reichardt\altaffilmark{\Melbourne},
T.~Schrabback\altaffilmark{\Bonn}
}

\email{Email: mcdonald@space.mit.edu}   


\begin{abstract}
We present a multi-wavelength analysis of the four most relaxed clusters in the South Pole Telescope 2500 deg$^2$ survey, which lie at $0.55 < z < 0.75$. This study, which utilizes new, deep data from the \emph{Chandra X-ray Observatory} and \emph{Hubble Space Telescope}, along with ground-based spectroscopy from Gemini and Magellan, improves significantly on previous studies in both depth and angular resolution, allowing us to directly compare to clusters at $z\sim0$. We find that the temperature, density, and entropy profiles of the intracluster medium (ICM) are very similar among the four clusters, and share similar shapes to clusters at $z\sim0$. 
Specifically, we find no evidence for deviations from self similarity in the temperature profile over the radial range 10\,kpc $< r <$ 1\,Mpc, implying that the processes responsible for preventing runaway cooling over the past $\gtrsim6$\,Gyr are, at least roughly, preserving self similarity.
We find typical metallicities of $\sim$0.3Z$_{\odot}$ in the bulk of the ICM, rising to $\sim$0.5Z$_{\odot}$ in the inner $\sim$100\,kpc, and reaching $\sim$1Z$_{\odot}$ at $r<10$\,kpc. This central excess is similar in magnitude to what is observed in the most relaxed clusters at $z\sim0$, suggesting that both the global metallicity and the central excess that we see in cool core clusters at $z\sim0$ were in place very early in the cluster lifetime and, specifically, that the central excess is not due to late-time enrichment by the central galaxy. 
Consistent with observations at $z\sim0$, we measure a diversity of stellar populations in the central brightest cluster galaxies of these four clusters, with star formation rates spanning a factor of $\sim$500, despite the similarity in cooling time, cooling rate, and central entropy. These data suggest that, while the details vary dramatically from system to system, runaway cooling has been broadly regulated in relaxed clusters over the past 6\,Gyr.

%
%


\end{abstract}

%

\section{Introduction}
\setcounter{footnote}{0}

Galaxy clusters that are dynamically relaxed -- defined based on either the dynamics and distributions of the member galaxies \citep[e.g.,][]{carlberg97b,wen13,old18} or the smoothness and symmetry of the X-ray emitting intracluster medium \citep[ICM;][]{mohr93,mohr95,buote95, jeltema05, nurgaliev13,rasia13, mantz15} -- tend to have very uniform properties. These relaxed clusters, also commonly referred to as ``cool core clusters'', have uniform density and temperature profiles \citep[e.g.,][]{vikhlinin06a,baldi12a,mantz16}, with the temperature dropping by a factor of $\sim$2 interior to $\sim$0.15R$_{500}$ \citep[e.g.,][]{vikhlinin06a}. 
They have metallicity profiles with peak values of $\sim$0.5--1.0 Z$_{\odot}$ at their centers \citep[e.g.,][]{degrandi01,baldi07,leccardi08b,mernier16,mantz17}, and reach a minimum of $\sim$0.2Z$_{\odot}$ outside of the core \citep[e.g.,][]{baldi12b,werner13,mcdonald16a,ezer17,mernier17,mantz17}.
These clusters tend to have a single massive galaxy at the cluster center \citep[e.g.,][]{haarsma10,rossetti16}, referred to as the brightest cluster galaxy (BCG) or the central cluster galaxy. This massive galaxy is almost always radio loud \citep[e.g.,][]{dunn06,sun09b} and is often forming stars at levels far lower than would be implied by predictions based on the cooling rate of the ICM \citep[e.g.,][]{odea08,mcdonald18a}. 

It is unclear when each of these properties of relaxed clusters were established. 
There is some evidence that the thermodynamic profiles have evolved self similarly since at least $z\sim1$ \citep{baldi12a,mantz16}, that the metallicity peaks were in place early \citep{ettori15,mcdonald16a,mantz17}, and that the central AGN were already radio loud $\sim$6 Gyr ago \citep{hlavacek12,hlavacek15}. However, much of our understanding of how galaxy clusters evolve is based on much shallower data compared to the depths that we routinely reach at $z\sim0$. Specifically, X-ray observations of galaxy clusters at $z\sim0$ have, on average, $>$100,000 counts, while those at $z\sim1$ have $\sim$2,000. In the optical, a typical ground-based observation of a galaxy cluster at $z\sim0$ has a physical resolution of $<$1\,kpc, while at $z\sim1$ the resolution is nearly an order of magnitude worse. This can complicate analyses and make it difficult to directly compare systems over a large redshift range.

In this work, we attempt to even the playing field, providing deep \emph{Chandra} and high resolution \emph{Hubble} observations in the X-ray and optical, respectively, to provide our first high-fidelity view of a sample of massive, relaxed clusters at $z\sim0.7$. The goal of this work is to establish the properties of the most relaxed clusters in a mass-selected sample of high-$z$ clusters, using data of similar quality to that obtained for low-$z$ clusters. Specifically, we focus on the properties of the cluster core (thermodynamics, metallicity) and the central galaxy (morphology, stellar populations). We defer an analysis of the dynamical state of these clusters and the properties of their central AGN to a companion paper. In \S2 we define the sample, which is drawn from the South Pole Telescope (SPT) 2500 deg$^2$ SPT-SZ survey \citep{bleem15}, and describe the acquisition, reduction, and analysis of the X-ray and optical data. In \S3 we present the results of this analysis, focusing on the thermodynamic profiles, the metallicity profiles, and the stellar populations of the central galaxy. In \S4 we discuss these results, focusing on understanding the connection between the ICM and the central galaxy, and on understanding the lack of evolution in the metallicity profile. We finish in \S5 with a summary of this work, and a look towards the future. Throughout this work, we assume H$_0 = 70$ km s$^{-1}$ Mpc$^{-1}$, $\Omega_M = 0.3$, and $\Omega_{\Lambda} = 0.7$. Unless otherwise stated, error bars represent 68\% confidence intervals.

\section{Data \& Analysis}

\subsection{Sample Selection}

This sample of four clusters was drawn from the larger SPT-Chandra sample of 100 galaxy clusters, which were selected via the Sunyaev Zel'dovich effect \citep{sunyaev72} by the SPT, and then followed up to a common depth ($\sim$2000 counts) with \emph{Chandra} \cite[see e.g., ][]{mcdonald13b,mcdonald17}. Of these 100 clusters, there are four that satisfy the conservative ``relaxed'' criterion, as described in \cite{mantz15}: SPT-CLJ0000-5748, SPT-CLJ2043-5035, SPT-CLJ2331-5051, and the Phoenix cluster (hereafter SPT-CLJ2344-4243). 
These clusters are all found to have a centrally-peaked surface brightness profile, with the peak centered on the large-scale X-ray centroid, and with isophotal ellipses that do not vary strongly in position angle \citep[see][for a further description of this selection]{mantz15}.
All four of these clusters also satisfy the relaxation criteria of \cite{nurgaliev16}, $A_{phot} < 0.2$, which (based on simulations) correspond to clusters that have not experienced a major merger in $\gtrsim$3\,Gyr.
The most relaxed of these systems, SPT-CLJ2344-4243 (Phoenix), has been the subject of numerous studies \citep[e.g.,][]{mcdonald12c, mcdonald13a, mcdonald14a, mcdonald15b}, and may be a rare example of runaway cooling in the ICM. 
Here, we present follow-up, multi-wavelength observations of these four relaxed clusters, which all have strong cool cores \citep{mcdonald13a}, evidence for strong radio-mode AGN feedback \citep{hlavacek15}, and star-forming central galaxies \citep{mcdonald16b}, similar to their low-$z$ counterparts \citep[e.g.,][]{mcdonald18a}.
For each of these clusters, we have obtained deep \emph{Chandra} and \emph{Hubble Space Telescope} data, along with ground-based spectroscopy, in order to assess in greater detail the properties of the strongest cool cores as they were 6\,Gyr ago.

\begin{deluxetable*}{ccccccccccc}[htb]
\tabletypesize{\footnotesize} 
\tablecolumns{11}
\tablewidth{0pt}
\tablecaption{X-ray Properties of Relaxed SPT-Selected Clusters}
\tablehead{
\colhead{Cluster} & 
\colhead{$z$} &
\colhead{M$_{500,HE}$} &
\colhead{M$_{500,Y_X}$} &
\colhead{M$_{2500,HE}$} &
\colhead{R$_{2500}$} &
\colhead{\.M$_{cool}$} & 
\colhead{K$_{0}$} &
\colhead{$t_{cool,0}$} & 
\colhead{Z$_{0.0-0.1}$} & 
\colhead{Z$_{0.1-0.5}$} \\
\colhead{Name} & 
\colhead{ } &
\colhead{[10$^{14}$ M$_{\odot}$]} &
\colhead{[10$^{14}$ M$_{\odot}$]} &
\colhead{[10$^{14}$ M$_{\odot}$]} &
\colhead{[kpc]} &
\colhead{[M$_{\odot}$ yr$^{-1}$]} & 
\colhead{[keV cm$^2$]} &
\colhead{[Gyr]} & 
\colhead{[Z$_{\odot}$]} & 
\colhead{[Z$_{\odot}$]} 
}
%
 \startdata
SPT CLJ0000-5748  & 0.7019 &   9.7$_{-4.8}^{+5.9}$ & 4.1$_{-0.6}^{+0.7}$ & 2.1$_{-0.4}^{+0.5}$ & 408$_{-24}^{+29}$ & $  401 \pm 30$ & 11$_{-2}^{+3}$ & 0.21$_{-0.03}^{+0.03}$ & 0.58$_{-0.08}^{+0.09}$ & 0.27$_{-0.09}^{+0.09}$ \\
SPT CLJ2043-5035  & 0.7234 & 9.3$_{-3.3}^{+4.2}$ & 4.2$_{-0.2}^{+0.1}$ &1.5$_{-0.2}^{+1.2}$ & 360$_{-21}^{+81}$ & $  630 \pm 56$ & 12$_{-3}^{+3}$ & 0.21$_{-0.04}^{+0.03}$ & 0.44$_{-0.05}^{+0.05}$ & 0.28$_{-0.07}^{+0.07}$ \\
SPT CLJ2331-5051 & 0.5760 & 6.8$_{-1.7}^{+2.1}$ & 4.3$_{-0.4}^{+0.3}$ & 2.6$_{-0.5}^{+0.8}$ & 461$_{-30}^{+44}$ & $  294 \pm 24$ & 15$_{-5}^{+5}$ & 0.32$_{-0.08}^{+0.06}$ & 0.49$_{-0.08}^{+0.08}$ & 0.15$_{-0.06}^{+0.05}$\\
SPT CLJ2344-4243 & 0.5970 & 13.5$_{-2.7}^{+3.6}$& 14.3$_{-0.9}^{+0.8}$ & 6.3$_{-0.8}^{+0.8}$ & 613$_{-27}^{+26}$ & $2366 \pm 60$ & 16$_{-3}^{+2}$ & 0.18$_{-0.02}^{+0.01}$ & 0.47$_{-0.03}^{+0.04}$ & 0.39$_{-0.07}^{+0.06}$
\enddata
\tablecomments{Masses are calculated assuming hydrostatic equilibrium (HE) and the Y$_X$--M relation from \cite{vikhlinin09a}. Central quantities (K$_0$, $t_{cool,0}$) are measured at a radius of 5\,kpc. Metallicities are measured in annuli of 0.0--0.1R$_{500}$ and 0.1--0.5R$_{500}$, following \cite{mantz17}, where R$_{500}$ is based on the Y$_X$--M relation. A description of these parameters, and how they were derived, can be found in \S2.2.
}
\label{table:xray}
\end{deluxetable*}

\subsection{Chandra X-ray Data}

X-ray observations for each of the four clusters in our sample were initially obtained as part of the larger SPT-Chandra survey (OBSIDs: 9333, 9335, 13401, 13478; PIs: Garmire, Benson). These initial observations yielded $\sim$2000 counts per cluster, which was sufficient to determine their global metallicity \citep{mcdonald16a}, gas fraction \citep{chiu16, chiu18}, whether they were cool core \citep{mcdonald13b}, their dynamical state \citep{mantz15, nurgaliev16}, and to provide tentative detections of X-ray cavities in their core \citep{hlavacek15}. These clusters were then followed up with \emph{Chandra} to reach count levels of $\sim$10,000 (OBSIDs: 16135, 16545, 18238, 18239, 18240, 18241, 19695, 19697; PIs: McDonald, Hlavacek-Larrondo). To achieve these count levels, we required 218\,ks (SPT-CLJ0000-5748), 189\,ks (SPT-CLJ2043-5035), 151\,ks (SPT-CLJ2331-5051), and 131\,ks (SPT-CLJ2344-4243).

All \emph{Chandra} data were reduced using CIAO v4.9 and CALDB v4.7.7. For each cluster, we reprocess the data using \emph{chandra\_repro}, cleaning the ACIS background in ``very faint'' mode. Point sources were identified using an automated routine following a wavelet decomposition technique \citep{vikhlinin98}, and then visually inspected before masking. Each OBSID was filtered for flares using the ChIPS routine \emph{lc\_clean}, and blank sky background files were obtained using the \emph{blanksky} routine. Background files were renormalized in the 10--12 keV bandpass (at which energies the effective area of \emph{Chandra} is negligible) to match each observation. In addition to the blank-sky background, we extract an off-source spectrum for each observation at a physical distance of $>$3 Mpc from the cluster center, allowing us to better constrain the astrophysical background on an exposure-by-exposure basis. 

We extract X-ray spectra in concentric annuli centered on the X-ray peak, using two separate binnings, one fine and one coarse. The coarse binning has sufficient width to provide $\gtrsim$2000 counts per bin, allowing the measurement of spectroscopic quantities such as temperature and metallicity. These spectra are modeled in XSPEC v12.9.0\footnote{\textsc{APEC} normalizations have been corrected for a known bug which leads to underestimated densities by a factor of (1+$z$) (\url{https://heasarc.gsfc.nasa.gov/docs/xanadu/xspec/issues/archive/issues.12.9.0u.html}).} \citep{arnaud96} over the energy range 0.7--7.0 keV  using a combination of Galactic photoelectric absorption (\textsc{phabs}), an optically thin plasma to represent the ICM (\textsc{apec}), and two background components consisting of Galactic emission (\textsc{apec}, $kT=0.18$ keV, $Z=Z_{\odot}$, $z=0$) and unresolved point sources (\textsc{bremss}; $kT=40$ keV). The two background components are joint-fit to the on-source and off-source spectra, with their normalizations per unit area tied between regions.  When measuring spectroscopic temperature and metallicity for SPT-CLJ2344-4243, we mask the inner 2.5$^{\prime\prime}$, which is contaminated by a strong central point source.

We also extract spectra in finely-spaced annuli, starting at 1$^{\prime\prime}$ in width and growing as needed to be signal-dominated, for the purpose of measuring the emission measure profile. When modeling these spectra, we freeze the temperature and metallicity of the ICM to the interpolated values from the coarse temperature profile. This allows us to reduce the degrees of freedom in the fit, and constrain the density profile with much higher resolution. For the inner 2.5$^{\prime\prime}$ of SPT-CLJ2344-4243, which is heavily contaminated by a central point source, we consider only energies $<$2\,keV. These energies are free from emission from the highly-obscured (type-II) central QSO \citep{ueda13}, and this narrow energy band is sufficient to constrain a single free model parameter (normalization).
We convert from the \textsc{apec} normalization to emission measure using $\int{n_en_pdV}=N\times4\pi\times10^{14}\left[D_A(1+z)\right]^2$, where $N$ is the \textsc{apec} normalization and $D_A$ is the angular diameter distance to the cluster.

Emission measure profiles were fit by numerically integrating the three-dimensional density profile along the line of sight and over the width of each annulus, producing a projected profile. We assume that the three-dimensional profile is of the form described by \cite{vikhlinin06a}:

{\small
\begin{equation}
n_pn_e = n_0^2 \frac{(r/r_c)^{-\alpha}}{(1+r^2/r_c^2)^{3\beta - \alpha/2}} \frac{1}{(1+r^{\gamma}/r_s^{\gamma})^{\epsilon/\gamma}} + \frac{n_{0,2}^2}{(1+r^2/r_c^2)^{3\beta_2}}~,
\label{eq:dens}
\end{equation}
}

\noindent{}where we leave all parameters free except for $\gamma$, which is fixed to $\gamma=3$, following \cite{vikhlinin06a}. The projected profile is fit to the data using the MPFITFUN procedure in IDL. We fit 100 realizations of the data, where data points are allowed to vary between fits based on their uncertainties, which provides an uncertainty in the fit. To convert from $n_en_p$ to $n_e$, we assume $n_e = \sqrt{n_e^2} = \sqrt{1.199n_en_p}$, where $Z=n_e/n_p=1.199$ is the average nuclear mass for a plasma with 0.3Z$_{\odot}$ metallicity, assuming abundances from \cite{anders89}.

The temperature profiles, which have betwen 7--9 radial bins, are fit using the MPFITFUN procedure in IDL with a modified version of the model from \cite{vikhlinin06a}: 

{\small
\begin{equation}
T(r) = T_0 \frac{(r/r_{core})^{\alpha}+T_{min}/T_0}{(r/r_{core})^{\alpha}+1} \frac{1}{(1+(r/r_{out})^2)}~.
\label{eq:temp}
\end{equation}
}

\noindent{}This profile only has 5 free parameters, compared to the more general profile from \cite{vikhlinin06a}, which has 9. We project this three-dimensional temperature model along the line of sight, and over the width of bin, using our model density profile from above and assuming that

{\small
\begin{equation}
\left<T\right> = \frac{\int_V wTdV}{\int_V wdV}~,
\label{eq:proj}
\end{equation}
}

\noindent{}where

{\small
\begin{equation}
w=n_e^2T^{-0.75}~,
\end{equation}
}

\noindent{}following \cite{vikhlinin06b}. This projected model was fit to the data, again using MPFITFUN and bootstrapping over 500 realizations of the temperature profile to provide uncertainties on the model.

In Table \ref{table:xray}, we summarize some of the relevant X-ray properties for each cluster. We include the total mass M$_{\Delta}$ measured within R$_{\Delta}$, the radius within which the average enclosed density is $\Delta$ times the critical density, where $\Delta=500,2500$. These estimates are calculated from the pressure ($P\equiv n_ekT$) profile, assume hydrostatic equilibrium. The estimates of M$_{500}$ are poorly constrained due to the large uncertainties in the temperature model at $\gtrsim$0.5\,Mpc, so we also quote M$_{500}$ from the Y$_X$--M scaling relation \citep{vikhlinin09a} -- throughout this work, quoted R$_{500}$ values will be derived from this scaling relation rather than from hydrostatic masses. We also include in this table
the classical cooling rate \citep[\.M$_{cool} \equiv \frac{M_{g}(<r_{cool})}{t_{cool}}$, following][]{mcdonald18a}, the central entropy ($K\equiv kTn_e^{-2/3}$), the central cooling time ($t_{cool} \equiv \frac{3}{2}
\frac{(n_e+n_p)kT}{n_en_H\Lambda(kT,Z)}$), and the average metallicity of the ICM in the inner (0.0--0.1R$_{500}$) and outer (0.1--0.5R$_{500}$) parts of the cluster. 
With the exception of the metallicities, which come from the spectral fitting, these are all derived directly from the three-dimensional density and temperature profiles, described above.


\subsection{Optical Imaging}

SPT-CLJ0000-5748 and SPTCLJ2331-5051 were observed with HST/ACS as part of program GO 12246 (PI: Stubbs)  between Sep 29, 2011 and Nov 27, 2011 using a single central pointing in F814W and a \mbox{$2\times 2$} mosaic in F606W. Each pointing was observed for 1.92ks split into four exposures to facilitate cosmic ray removal. For these clusters we employ the reduction described in \cite{schrabback18}, which uses the \cite{massey14} algorithm for the pixel-level correction for the impact of charge transfer inefficiency, CALACS for basic image reductions, scripts from \cite{schrabback10} for the image registration and weight optimization, and MultiDrizzle \citep{koekemoer03} for the cosmic ray removal and stacking.

We reduced HST/ACS observations of SPT-CLJ2043-5035 using the same pipeline, combining \mbox{$2\times 2$} mosaics obtained between Oct 20 and 25, 2015 in both F814W (1.96ks per pointing) and F606W (1.93ks per pointing) via program GO 14352 (PI: Hlavacek-Larrondo) with central single-pointing F606W observations (1.44ks) obtained on May 24, 2014 via program SNAP 13412 (PI: Schrabback).
%
%
%
%
%
 For all three clusters, the two available filters span the 4000\AA\ break, providing a spatially-resolved view of both the old and young stellar populations in the central brightest cluster galaxy (BCG) for each system.
 
SPT-CLJ2344-4243 was observed with WFC3-UVIS as part of the HST program GO 13102 (PI: McDonald) at F625W and F814W. Details of these data and their analysis are presented in \cite{mcdonald13a}. These data are shallower than the other three clusters, but are sufficient for the purposes of this study, particularly because of the relative brightness of the central galaxy compared to other clusters.

\begin{figure}[htb]
\centering
\includegraphics[width=0.48\textwidth, trim=0.4cm 0cm 1.2cm 1.0cm,clip]{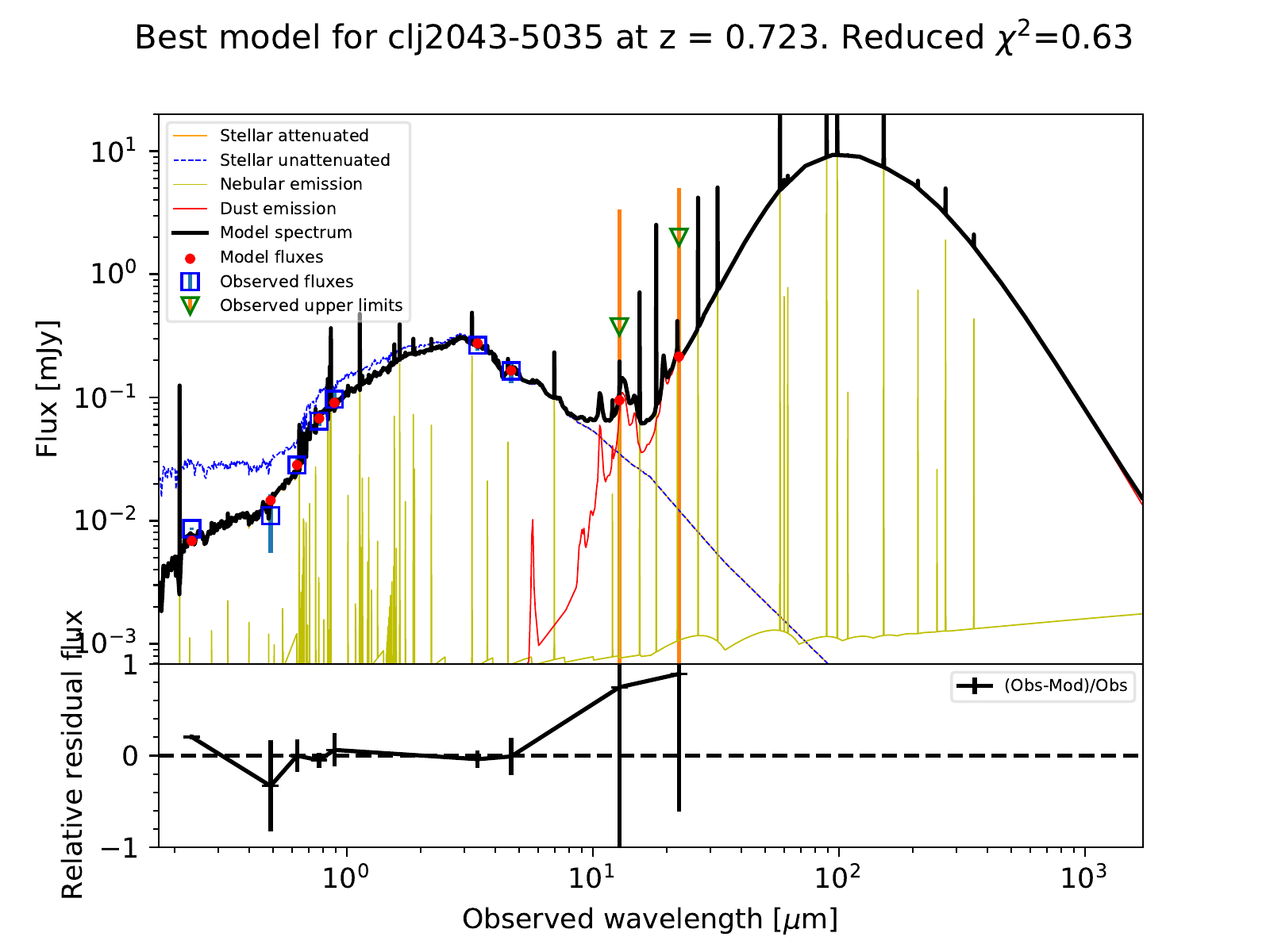}
\caption{Rest-frame UV-through-IR spectral energy distributions (SEDs) for SPT-CLJ2043-5035. This SED, which was presented in \cite{mcdonald16b}, has been fit using the CIGALE code, as described in \S2.3. For all four BCGs in this paper, the CIGALE fits provide constraints on the total stellar mass of the old and young populations, along with the amount of intrinsic extinction due to dust within the system. 
The best fit model is shown in black, but we note that we consider the range of allowable models in this analysis in order to fold the uncertainty on the extinction into estimates of the (for example) intrinsic [O\,\textsc{ii}] luminosity.
}
\label{fig:bcg_seds}
\end{figure}

\begin{figure*}[htb]
\centering
\includegraphics[width=0.9\textwidth]{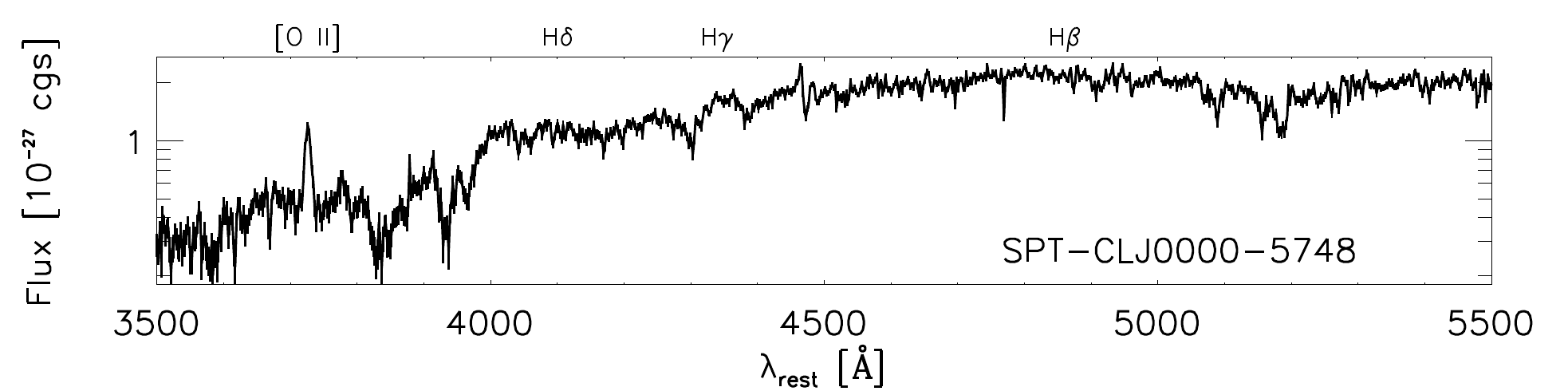}\\ 
\includegraphics[width=0.9\textwidth]{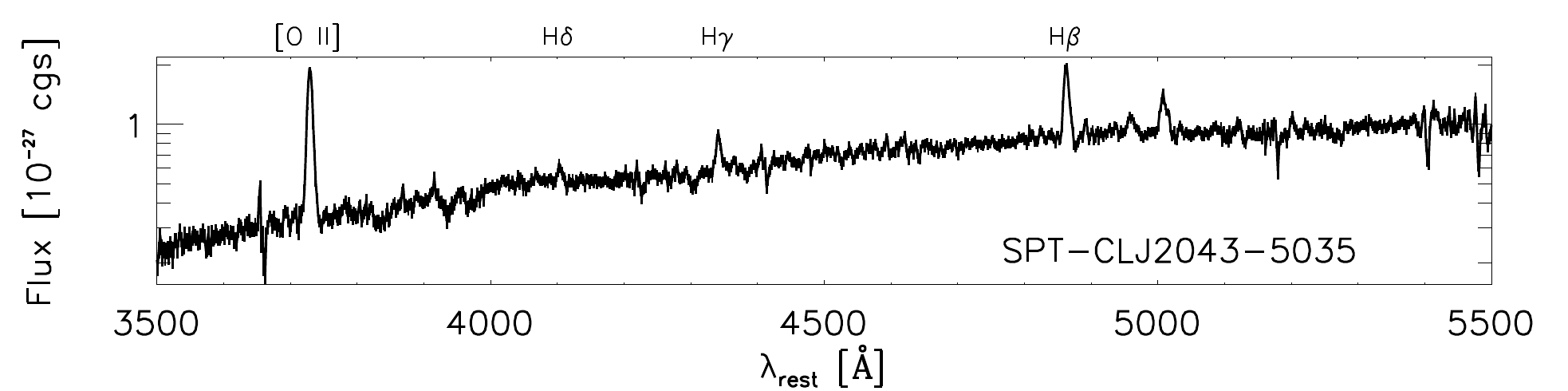}\\
\includegraphics[width=0.9\textwidth]{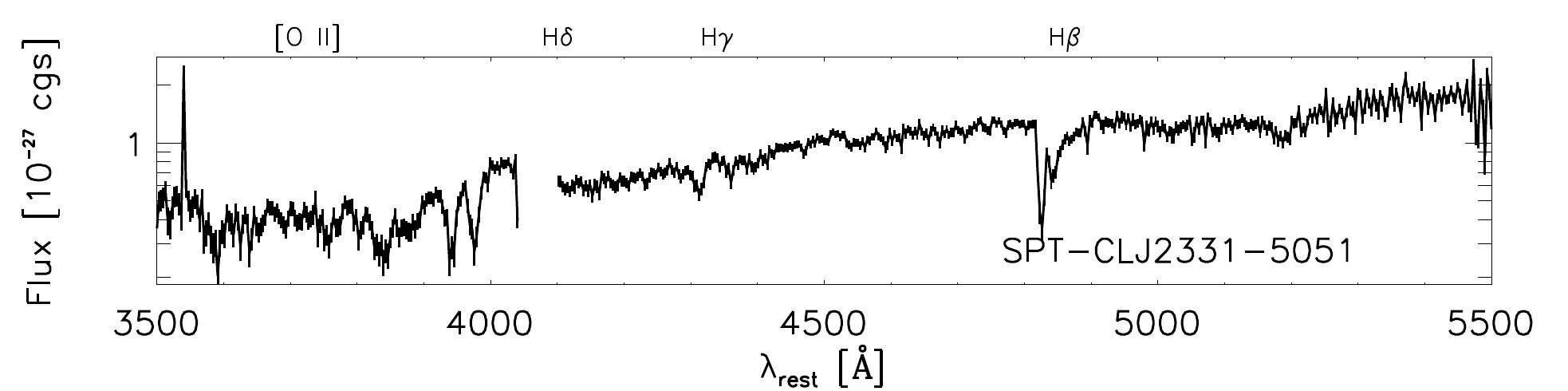}\\
\includegraphics[width=0.9\textwidth]{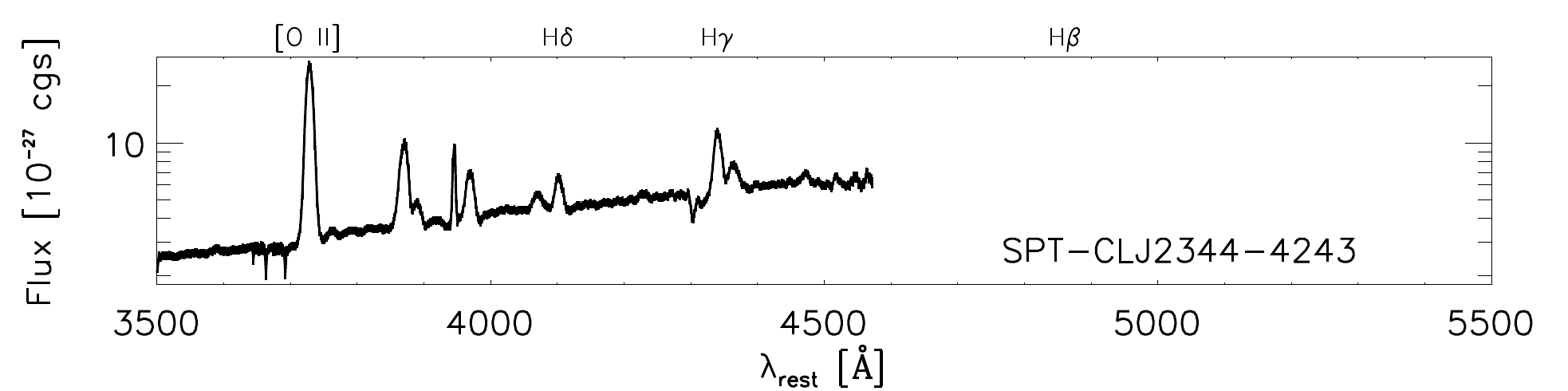}
\caption{Optical spectra for the central, brightest cluster galaxy in each of our four clusters. These systems show a variety of features, including strong [O\,\textsc{ii}] emission (SPT-CLJ2043-5035, SPT-CLJ2344-4243), weak [O\,\textsc{ii}] emission (SPT-CLJ000-5748), strong 4000\AA\ breaks (SPT-CLJ0000-5748, SPT-CLJ2331-5051), and a variety of absorption lines (SPT-CLJ0000-5748, SPT-CLJ2331-5051). For the analysis presented in this work, we consider only the brightness and width of the [O\,\textsc{ii}]$\lambda\lambda$3726,3729 doublet relative to the nearby continuum. We note that the missing data in the spectrum for SPT-CLJ2331-5051 is due to a chip gap, and the bright narrow line at $\sim$3500\AA\ is a residual sky line. For SPT-CLJ2344-4243 the choice of grating led to sensitivity at $\lambda_{obs} < 7350$\AA\ only.}
\label{fig:spectra}
\end{figure*}

Ground- and space-based broadband observations spanning rest-frame 0.1--13$\mu$m were acquired for each BCG from \cite{mcdonald16b}. Details of the data acquisition are presented therein.
We use CIGALE\footnote{\url{https://cigale.lam.fr/}} \citep{cigale} to estimate the SED-based stellar mass and intrinsic extinction due to dust for each BCG. The combination of rest-frame UV data (observed $u$ band) and mid-IR data from WISE provides strong constraints on the total stellar mass and the amount of intrinsic extinction -- if the UV light is suppressed by extinction, we expect strong mid-IR emission; if there is no mid-IR emission, we expect the UV emission to be relatively unextincted. For all four BCGs we include an old and young stellar population, attenuation and emission due to dust, and nebular lines due to warm, ionized gas. 
For these fits, we assume \cite{calzetti94} extinction, dust emission described by \citep{dale14}, a \cite{salpeter55} initial mass function, and a grid of old (4, 8, 12 Gyr) populations, starburst delay times (10, 20, 40, 80, 160, 320, 640 Myr), e-folding times (50, 250, 500, 1000, 2000, 4000, 6000, 8000 Myr), metallicities (0.004, 0.02, 0.05), and reddening ($E(B-V)$ $=$ 0.0, 0.2, 0.4, 0.6 mag).
We include an AGN component for the BCG in SPT-CLJ2344-4243 based on photometric and spectroscopic observations in the UV-optical-IR \citep[see e.g.,][]{mcdonald15b} -- for the other three clusters there is no supporting evidence for an optical-IR bright nucleus, and the inclusion of an AGN weakens the constraints. We note that these estimates are based on aperture photometry (to avoid source confusion), with no attempt made to model the contribution to the total luminosity from large radii (i.e., intracluster light).

The results of this analysis for SPT-CLJ2043-5035, as an example, is shown in Figure \ref{fig:bcg_seds}, and for all four systems in Table \ref{table:bcg}. As is shown in Figure \ref{fig:bcg_seds}, the data at both $u$-band and WISE W3 and W4 (upper limits) bands provide joint constraints on the total (obscured and unobscured) mass of the youngest stellar populations.

\begin{deluxetable*}{ccccccccc}[htb]
\tabletypesize{\footnotesize} 
\tablecolumns{9}
\tablewidth{0pt}
\tablecaption{Properties of Central Brightest Cluster Galaxy}
\tablehead{
\colhead{Cluster} & 
\colhead{$\alpha_{BCG}$} & 
\colhead{$\delta_{BCG}$} & 
\colhead{$\Delta_{X-BCG}$} & 
\colhead{M$_{*,BCG}$} &
\colhead{L$_{[O II], BCG}$} & 
\colhead{E(B-V)$_{SED}$} &
\colhead{SFR$_{[O II], BCG}$} & 
\colhead{sSFR$_{BCG}$} \\
%
\colhead{Name} & 
\colhead{[$^{\circ}$]} &  
\colhead{[$^{\circ}$]} & 
\colhead{[${\prime\prime}$/kpc]} & 
\colhead{[10$^{11}$ M$_{\odot}$]} &
\colhead{[10$^{41}$ erg s$^{-1}$]} & 
\colhead{} &
\colhead{[M$_{\odot}$ yr$^{-1}$]} &
\colhead{[Gyr$^{-1}$]} 
}
%
 \startdata
SPT CLJ0000-5748 &   0.2503 & -57.8093 & 0.7/5.2 & $12.4_{- 0.7}^{+ 1.3}$ & $ 15.1 \pm   0.9$ & $0.17_{-0.13}^{+0.18}$ & $  17.8_{-   6.2}^{+  13.7}$ & $0.014_{-0.005}^{+0.011}$ \\ 
SPT CLJ2043-5035 & 310.8233 & -50.5923 & 1.5/10.7 & $ 4.1_{- 1.3}^{+ 0.0}$ & $ 34.8 \pm   0.9$ & $0.11_{-0.04}^{+0.06}$ & $  33.1_{-   4.6}^{+   7.1}$ & $0.090_{-0.017}^{+0.037}$ \\ 
SPT CLJ2331-5051 & 352.9631 & -50.8645 & 1.6/10.6 & $ 9.3_{- 2.4}^{+ 3.1}$ & $  1.5 \pm   1.2$ & $0.17_{-0.13}^{+0.21}$ & $   1.8_{-   1.4}^{+   2.4}$ & $0.002_{-0.001}^{+0.003}$ \\ 
SPT CLJ2344-4243 & 356.1831 & -42.7201 & $<$0.5/$<$3.3 & $14.5_{- 0.6}^{+ 0.8}$ & $339.6 \pm  10.6$ & $0.39_{-0.05}^{+0.04}$ & $ 809.7_{- 116.4}^{+ 110.0}$ & $0.554_{-0.082}^{+0.081}$ 
\enddata
\tablecomments{The BCG separation ($\Delta_{X-BCG}$) is the projected distance between the brightness peak of the galaxy identified as the BCG, and the soft X-ray peak. Due to the highly clumpy morphology of the BCG in SPT-CLJ2344-4243, which makes the center challenging to identify, the quoted offset is considered an upper limit. Quoted L$_{[OII]}$ values are uncorrected for intrinsic extinction. Star formation rates (SFRs) and specific star formation rates (sSFR) are derived from the [O\,\textsc{ii}] flux and include combined uncertainty in the [O\,\textsc{ii}] flux, the intrinsic extinction (E(B-V)), and the stellar mass. Stellar masses and intrinsic extinction are derived based on the SED fit (see \S2.3) -- the former do not contain the extended cD envelope. 
}
\label{table:bcg}
\end{deluxetable*}

\subsection{Gemini/Magellan Optical Spectroscopy}

Long-slit optical spectra for the central BCG in SPT-CLJ0000--5748, SPT-CLJ2043--5035, and SPT-CLJ2344-4243 were obtained as a part of dedicated multi-slit observing campaigns to measure the redshifts of dozens of cluster member galaxies. The BCGs in SPT-CLJ0000--5748 and SPT-CLJ2344-4243 were observed with the Gemini Multi-Object Spectrograph \citep[GMOS;][]{hook04} on the Gemini South telescope in September 2010 (GS-2009B-Q-16) and November 2011 (GS-2011A-C-3), respectively. The BCG in SPT-CLJ2043--5035 was observed with the Focal Reducer and low dispersion Spectrograph \citep[FORS2;]{appenzeller98} on the Very Large Telescope (VLT) in August 2011 as a part of ESO program 087.A-0843. 
The BCG in SPT-CLJ2331--5051 was observed with the Inamori-Magellan Areal Camera \& Spectrograph \citep[IMACS;][]{imacs} on the Magellan-I (Baade) telescope in December 2017 in long-slit mode as part of a dedicated program targeting BCGs. Spectra for all of these BCGs are relatively low--resolution ($\lambda / \Delta \lambda \sim 400-1000$) and cover wavelength ranges spanning most of the red side of the optical ($\Delta \lambda \sim5000-10000$\AA), corresponding to rest-frame blue ($\Delta \lambda_{rest} \sim3000-6000$\AA), with the exception of SPT-CLJ2344-4243 for which the choice of grating meant that the data span rest-frame 3500--4500\AA. We reduced the data using standard IRAF routines, with the GMOS, FORS2, and IMACS data making use of IRAF packages provided by Gemini, ESO, and Magellan, respectively. For further details of the observing strategy and data reduction methodology, we refer the reader to \cite{ruel14} and \cite{bayliss16}, in which most of these data were originally presented.

Some of the optical spectra used in this work are not flux calibrated using spectrophotometric standards, as they were obtained with the intention of measuring redshifts, rather than fluxes. We use the 0.1--13$\mu$m SEDs provided in \cite{mcdonald16b} to roughly estimate the flux calibration at 4 points ($g$,$r$,$i$,$z$) by convolving the uncalibrated spectrum with the filter bandpasses. This flux calibration is applied to the data, yielding the spectra shown in Figure \ref{fig:spectra}. These spectra are not suitable for full spectral modeling, due to the fact that their shapes are poorly constrained by only four data points. Locally, the calibration should not vary greatly, so relative, local measurements such as the strength of the 4000\AA\ break and equivalent width measurements of individual lines are relatively unaffected.

We use SED modeling of the broad-band, flux-calibrated photometry described in \S2.3 to estimate the flux at rest-frame 3727\AA, which we then combine with local spectroscopic measurements of the [O\,\textsc{ii}] equivalent width to estimate the calibrated flux of the [O\,\textsc{ii}] emission line doublet. 
We correct for intrinsic absorption due to dust using the measured attenuation from the CIGALE SED model, along with the uncertainty in this measurement (see Table \ref{table:bcg}).
When converting from the extinction-corrected [O\,\textsc{ii}] luminosity to star formation rate, we follow the prescriptions described in \cite{kewley04}. Quantities extracted from these photometric and spectroscopic analyses are provided in Table \ref{table:bcg} for each BCG.


\section{Results}

\subsection{Thermodynamic Profiles and Central Properties}

\begin{figure*}[htb]
\centering
\includegraphics[width=0.99\textwidth]{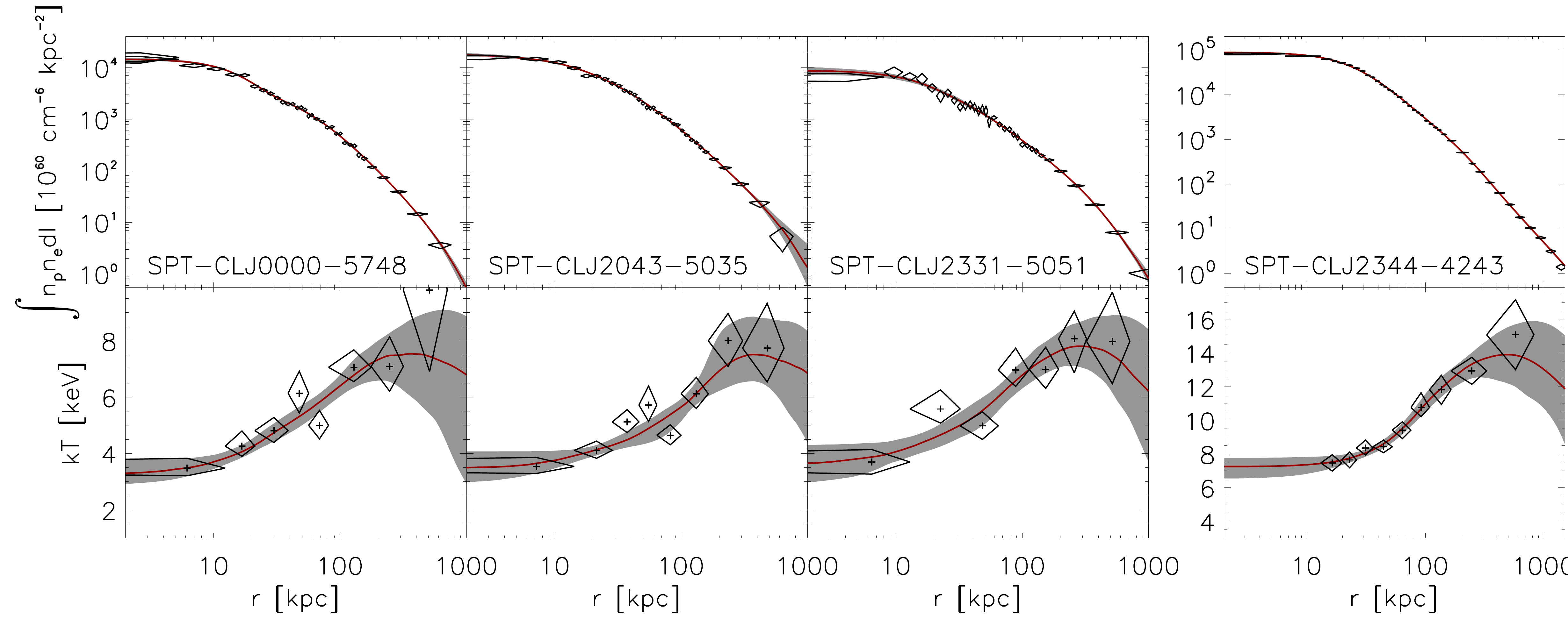}
\caption{Thermodynamic profiles for all four clusters, as described in \S2.2. We note that, since SPT-CLJ2344-4243 has a factor of $\sim$2 higher mass than the other three systems, we have adjusted the plotting limits for the temperature and density profiles.
\emph{Upper row:} These panels show the emission measure profile for each of the clusters, with sampling chosen to produce a high signal-to-noise, well-sampled profile. This projected profile is fit with a three-dimensional model (see Equation \ref{eq:dens}), which has been projected along the line of sight and along the radial direction within each bin. All four clusters exhibit a central overdensity, indicative of a cool core. \emph{Lower row:} These panels show the projected temperature profile, with a coarser sampling that reflects the added complexity of constraining the spectroscopic temperature and metallicity. These profiles have been fit with a three-dimensional model (see Equation \ref{eq:temp}) which has been projected along the line of sight following Equation \ref{eq:proj}. All four clusters have similar temperature profiles, reaching a maximum projected temperature at $\sim$300\,kpc that is $\sim$2$\times$ higher than the minimum, projected temperature measured at $\sim$10\,kpc.}
\label{fig:xray_profs}
\end{figure*}

For each cluster, we have measured the emission measure and projected temperature profile from X-ray spectra, as described in \S2.2. We model these profiles by projecting a three-dimensional model onto two dimensions, and then fitting the projected model to the data. We can then back out the analytic form of the three-dimensional temperature and density profiles, which can be used to infer the three dimensional entropy, cooling time, and pressure (which is used to determine the hydrostatic mass).

In Figure \ref{fig:xray_profs}, we show the results of this analysis. In the upper and lower panels, we show the projected emission measure and projected temperature profiles, respectively, along with the best fit models and the 1$\sigma$ uncertainty in this model. In all four clusters, the density profile is strongly peaked in the center, indicating the presence of a cool core. This is unsurprising -- our sample was selected to contain the most relaxed, strongest cool cores in the full SPT-Chandra sample of 100 clusters. All four of these clusters satisfy density-based criteria for cool cores \citep[see][]{semler12,mcdonald13b}, including the cuspiness \citep[$\alpha \equiv \frac{d\log\rho_g}{d\log r}\rvert_{r=0.04R_{500}} > 0.7$;][]{vikhlinin07} and concentration \citep[$C_{SB} \equiv \frac{F_X (r<40 kpc)}{F_X (r<400 kpc)} > 0.155$;][]{santos08}.

These density peaks are coincident with a significant drop in temperature at the cluster center.  We find three-dimensional central temperature drops ($T_{min}/T_0$, see equation 2) of 0.17$_{-0.07}^{+0.12}$, 0.23$_{-0.10}^{+0.19}$, 0.15$_{-0.07}^{+0.14}$, and 0.41$_{-0.15}^{+0.24}$ for SPT-CLJ0000-5748, SPT-CLJ2043-5035, SPT-CLJ2331-5051, and SPT-CLJ2344-4243, respectively. For comparison, \cite{vikhlinin06a} find $T_{min}/T_0$ ranging from 0.1--0.4 for the 5 most massive clusters in their sample of relaxed, low-$z$ clusters. We find cool core sizes, as measured from the temperature profile ($r_{core}$, see equation 2), of 135--250\,kpc for the four SPT clusters, compared to typical values of 30--214\,kpc for the most 5 massive clusters from \cite{vikhlinin06a}. Relative to the measured values of R$_{2500}$, these core radii are, on average, larger for SPT clusters ($\sim$40\% R$_{2500}$) compared to those from \cite{vikhlinin06a} ($\sim$20\% R$_{2500}$), however both samples are too small to draw definitive conclusions.
Overall, it appears that the temperature profiles in these four relaxed clusters at $z\sim0.7$ share similar shapes to those at $z\sim0$.

\begin{figure}[htb]
\centering
\includegraphics[width=0.49\textwidth]{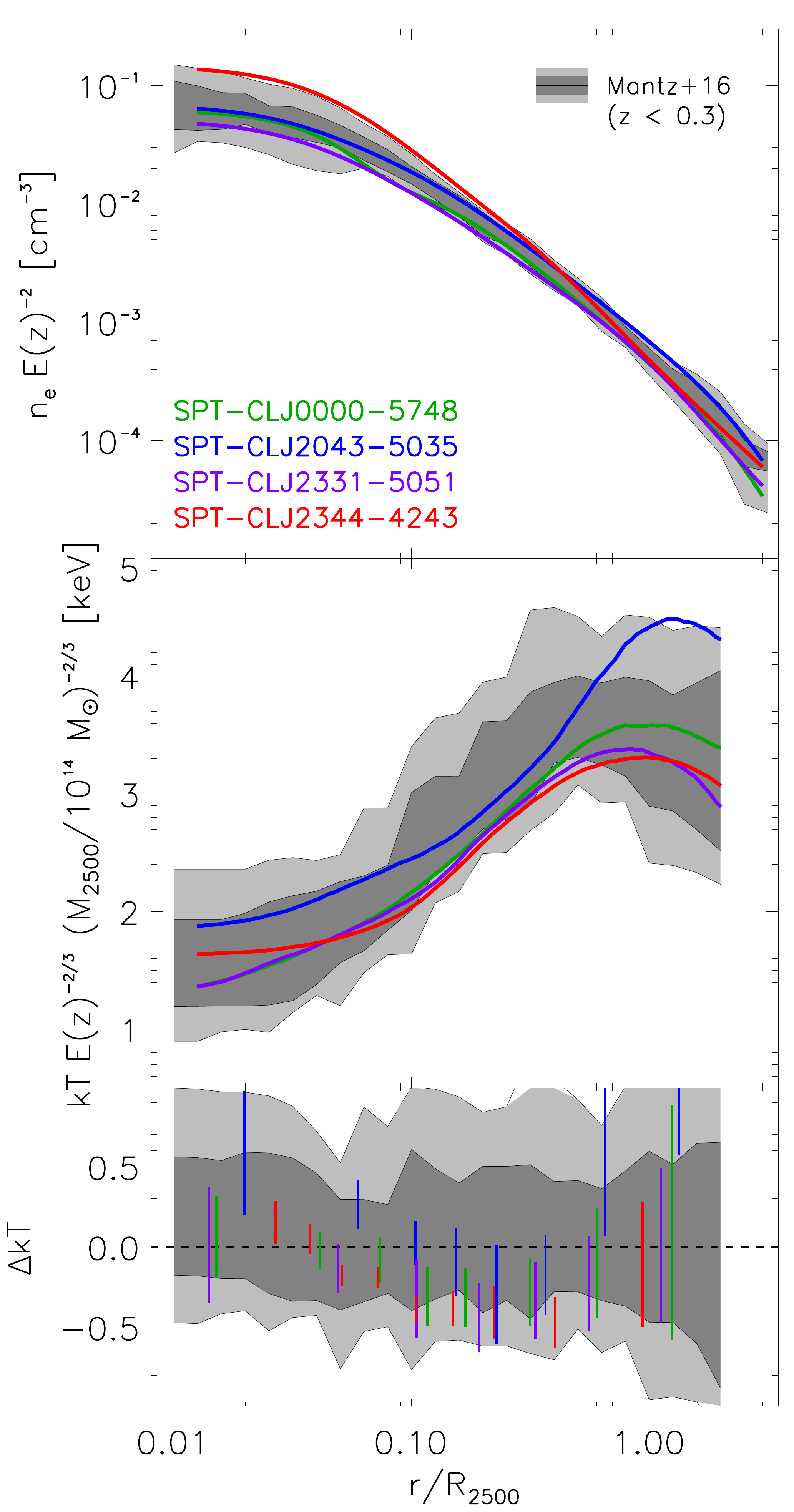}
\caption{In all three panels, we compare the three-dimensional thermodynamic profiles for our four relaxed clusters to those of a similarly-selected sample of relaxed clusters at $z<0.3$ from \cite{mantz16}. The latter are depicted as gray bands, with dark and light bands representing the 1$\sigma$ and 2$\sigma$ scatter. All profiles have been scaled according to the \cite{kaiser86} self-similar model, and are plotted as a function of scaled radius (R$_{2500}$). In the lower panel, we show the residual between individual clusters in our sample and the average relaxed cluster at $z<0.3$, demonstrating that, given the measurement uncertainty, all four clusters are consistent with self similar evolution over the past 6\,Gyr.}
\label{fig:selfsim}
\end{figure}

To further investigate the evolution of the three-dimensional density and temperature profiles, we compare to a similarly-selected sample of relaxed clusters at $z<0.3$ from \cite{mantz16}. The four systems in this work satisfy the same conservative relaxation criteria (namely that they are centrally peaked, azimuthally symmetric, and well-aligned) as the primarily X-ray-selected $z<0.3$ clusters described in \cite{mantz15}.
 In Figure \ref{fig:selfsim} we show the individual three-dimensional profiles, scaled according to the \cite{kaiser86} self-similar model and as a function of scaled radius (R$_{2500}$). For both the density and temperature profiles, the SPT clusters lie within the 1$\sigma$ loci defined by the $z<0.3$ relaxed cluster sample \citep{mantz16}, implying that their evolution is well-described by a self similar model at all radii. The only exception to this is the core of the Phoenix cluster (SPT-CLJ2344-4243), which is the strongest known cool core and, thus, traces the high-density edge of the 2$\sigma$ locus.
The four SPT clusters appear to have shallower inner slopes in the temperature profiles (reaching maxima at larger radii), consistent with the overall higher core radii mentioned above. 
We note that this slower rise of the temperature profile is consistent with the picture presented in \cite{mcdonald17}, where the cool core is a fixed physical size (corresponding to the ``reach'' of the central AGN) and the bulk of the cluster is evolving self similarly, leading to a decreasing ratio of the cool core to cluster size (R$_{500}$) with decreasing redshift.
However, we show in the lower panel of Figure \ref{fig:selfsim} that this is not statistically significant -- all four clusters lie within the 1$\sigma$ scatter for low-$z$ clusters when we consider uncertainties on our temperature measurements.
Further, we note that a $<$5\% systematic offset between our estimates of R$_{2500}$ and those of \cite{mantz16} -- which is a completely realistic offset between two independent analyses and on par with the statistical uncertainty in R$_{2500}$ -- is sufficient to remove this slight discrepancy.
Given these points, we would require significantly more than 4 clusters to claim any deviation from self similarity in the temperature profiles of relaxed clusters.

\begin{figure}[t]
\centering
\includegraphics[width=0.49\textwidth]{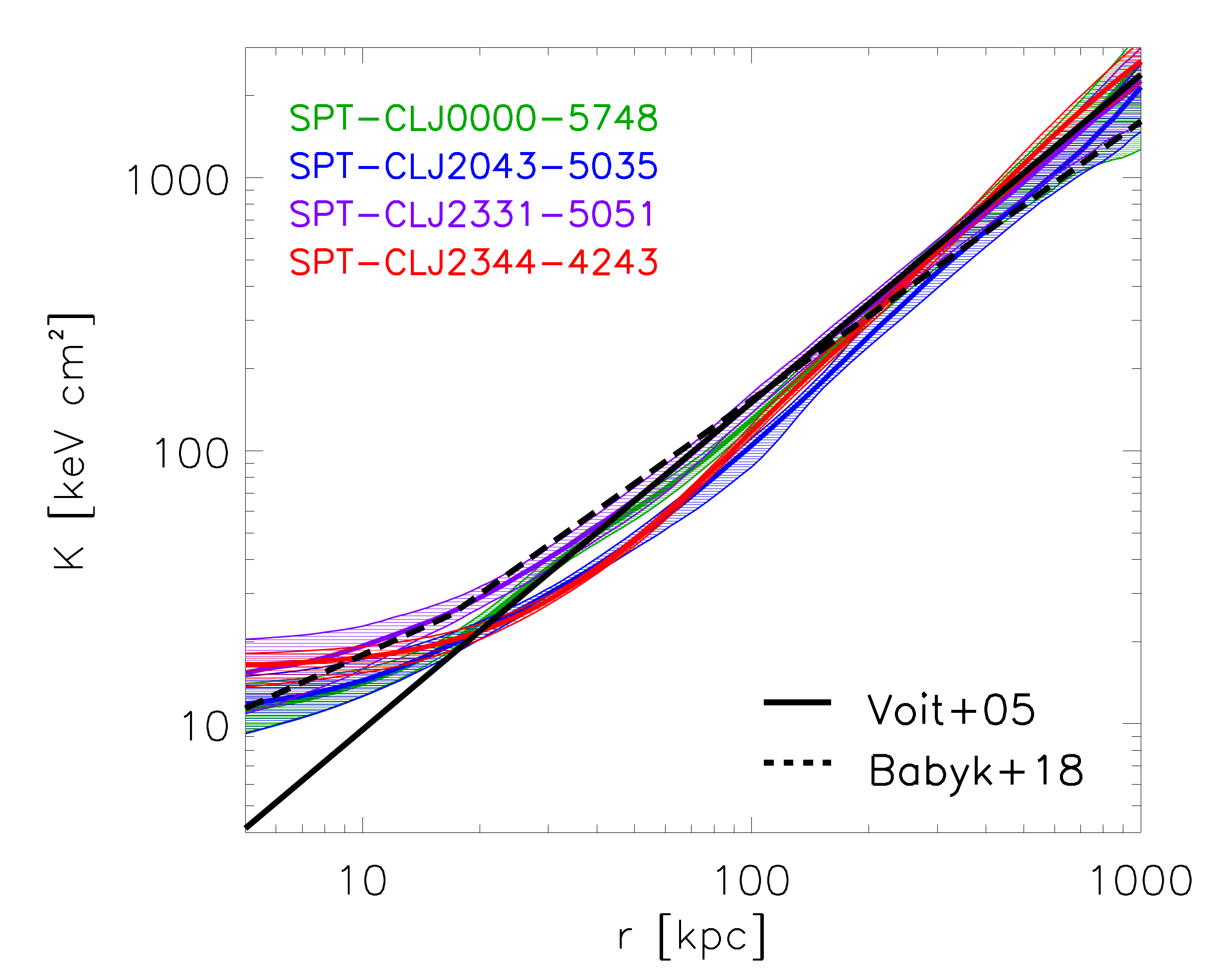}
\caption{The three-dimensional entropy profile is computed combining the model three-dimensional temperature and density profiles ($K\equiv kTn_e^{-2/3}$). These profiles all agree well with one another, and with those from the literature, at large and small radii.  We see a weak separation between ``high-entropy'' (SPT-CLJ0000-5748, SPT-CLJ2331-5051) and ``low-entropy'' (SPT-CLJ2043-5035, SPT-CLJ2344-4243) systems over the radial range 25--100\,kpc, with the two groups merging at smaller and larger radii.}
\label{fig:entropy}
\end{figure}


All four clusters have very similar density and temperature profiles, which naturally leads to very similar entropy ($K \equiv kT n_e^{-2/3}$) profile, as we show in Figure \ref{fig:entropy}. At large radii, these profiles follow the expectation if gravity is the only relevant physics \citep[$K \propto r^{1.2}$;][]{voit05}.  We see no evidence for a flattening of the entropy profile at radii near R$_{500}$ ($\sim$1\,Mpc), as was seen for more distant ($z>0.6$) clusters in \cite{mcdonald14c}. 
At small radii ($r \lesssim 30$ kpc),we detect a significant entropy excess above the gravity-only prediction, leading to a shallower slope for all three clusters. This slope change is consistent with what is observed at $z\sim0$ \citep{cavagnolo09,panagoulia14,babyk18}, and is indicative of baryonic physics (i.e., cooling, feedback) beginning to play a significant role. Over the full radial range considered, the profiles are consistent (at the 1--2$\sigma$ level) with the universal entropy profile for cool core clusters from \cite{babyk18}. 
There is marginal ($\sim$2$\sigma$) evidence for a dearth of entropy at $\sim$30--100\,kpc in SPT-CLJ2043-5035 and SPT-CLJ2344-4243, with these two profiles falling below both the \cite{voit05} and \cite{babyk18} profiles at the $\sim$2$\sigma$ level over this radial range. At radii smaller and larger than this intermediate region, the four profiles are statistically indistinguishable from one another. 

We measure the central entropy for each cluster at a radius of 5\,kpc, to avoid interpolating far beyond where our data can constrain. We find K$_0$ = 11, 12, 15, and 16 keV cm$^2$ for SPT-CLJ0000-5748, SPT-CLJ2043-5035, SPT-CLJ2331-5051, and SPT-CLJ2344-4243, respectively. These are all within 1$\sigma$ of the values quoted in \cite{mcdonald13b}, where we were unable to constrain the temperature profile due to a lack of counts, and instead  assumed a universal temperature profile with a single free parameter. This demonstrates that, at least for the most relaxed clusters, such an approach is valid. The small scatter in central entropies for these systems suggests a relatively gentle feedback cycle -- periods of runaway cooling and/or powerful AGN outbursts would act to increase the scatter in the central entropy for a sample of relaxed clusters.
All four of these cluster cores lie below the entropy threshold for multiphase gas \citep[K$_0 < 30$ keV cm$^2$;][]{cavagnolo08}, implying that the central galaxies in these clusters ought to have strong H$\alpha$ emission and other signatures of star formation -- we will return to this point in \S3.3 and in the discussion. 

\subsection{Metallicity Profiles and Central Metallicity Peak}

\begin{figure*}[htb]
\centering
\includegraphics[width=0.99\textwidth]{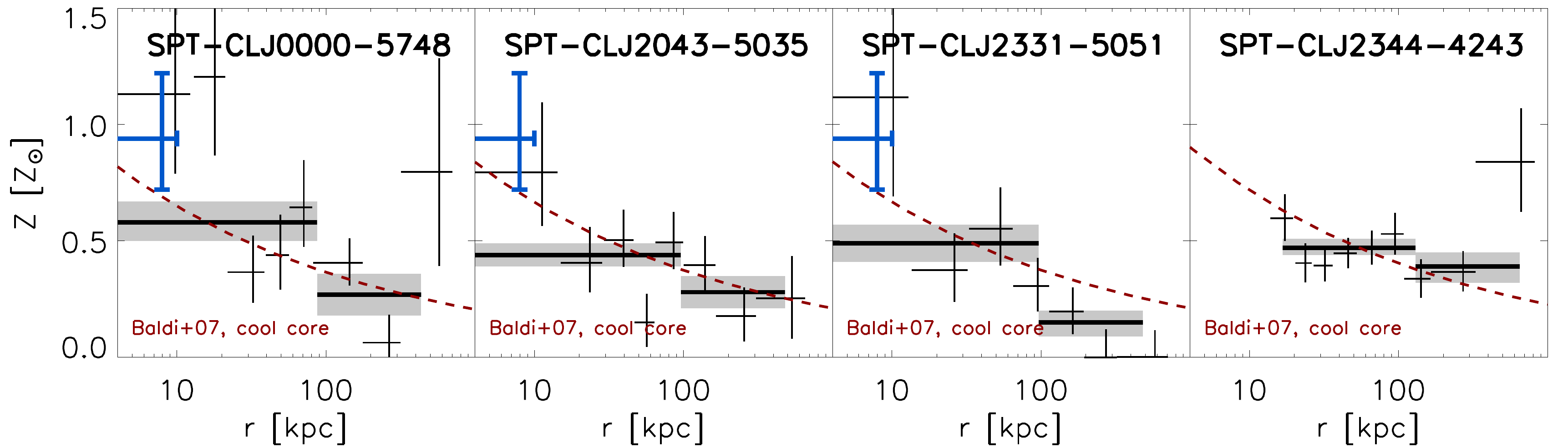}
\caption{ICM metallicity profiles for the four clusters in our sample. Black crosses show the metallicity measured with the same sampling as the temperature profile (see Figure \ref{fig:xray_profs}). We note that the inner 2.5$^{\prime\prime}$ for SPT-CLJ2344-4243 has been masked due to contamination by a bright central X-ray point source with an extremely high equivalent width Fe K$\alpha$ emission line. Thick black lines with grey boxes show the metallicity and its uncertainty measured in annuli of $0 < r < 0.1R_{500}$ and $0.1R_{500} < r < 0.5R_{500}$, following \cite{mantz17}. Blue points show the joint constraint on the metallicity in the inner 10\,kpc from SPT-CLJ0000-5748, SPT-CLJ2043-5035, and SPT-CLJ2331-5051, assuming all three clusters have a shared temperature and metallicity (see \S3.2 for more details). We compare these profiles to the average profile for low-$z$ cool core clusters \citep[red dashed line;][]{baldi07}, finding that the profiles agree well at all radii, with weak ($\sim$1$\sigma$) evidence for a slightly higher than average central ($r<10$\,kpc) metallicity in the high-$z$ clusters.}
\label{fig:metallicity}
\end{figure*}


In Figure \ref{fig:metallicity} we show the metallicity profiles for each cluster. We measure the metallicity in each bin for which we measure temperature (see \S2.2), and separately in bins of 0.0--0.1R$_{500}$ and 0.1--0.5R$_{500}$, following \cite{mantz17}. The latter measurements are quoted in Table \ref{table:xray}. We assume solar abundances from \cite{anders89} to be consistent with the bulk of the literature, but note that our metallicities can be roughly converted to those based on solar abundances from \cite{asplund09} by multiplying by a factor of 1.4. While not perfect, this multiplicative factor is accurate to better than our measurement uncertainties.


Figure \ref{fig:metallicity} shows steeply rising metallicity profiles for three of the four clusters in our sample. Outside of the innermost bin ($ r\gtrsim 15$\,kpc) these metallicity profiles are all consistent at the 1$\sigma$ level with the average profile for massive cool core clusters at $z\sim0$ \citep[][not shown]{degrandi01} and from $z=0.1-0.3$ \citep{baldi07}. In the innermost bin ($ r\lesssim 15$\,kpc), we find Z $\sim$ Z$_{\odot}$ in all three systems for which such a measurement is possible, with large uncertainties. To test the statistical significance of these highly-enriched central regions, we perform a somewhat unorthodox test. In the previous section, we have established how remarkably similar the thermodynamic profiles of these clusters are: in the inner $\sim$10\,kpc, all three of the lower-mass clusters (excluding SPT-CLJ2344-4243) have consistent temperatures, emission measures, entropies, and cooling times to within the 1$\sigma$ uncertainties. For this reason, we feel comfortable combining these three systems into a single stacked cluster in order to improve the constraints on the metallicity. We have extracted spectra in the inner 10\,kpc for all three clusters and fit them jointly with an APEC model for which we tie the metallicity and temperature, fix the column density and redshift to the nominal values, and allow the normalization (which is a function of distance) to vary. We find a combined central temperature of $3.67_{-0.23}^{+0.26}$ keV (consistent with Figure \ref{fig:xray_profs}) and a combined metallicity of $0.94_{-0.22}^{+0.28}$ Z$_{\odot}$ in the inner 10\,kpc. These data points are shown in Figure \ref{fig:metallicity} for comparison to the literature and the individual profiles. This central metallicity, which is a factor of $\sim$4$\times$ higher than that measured in the bulk (0.1--0.5R$_{500}$) of the four clusters, is higher than that measured at the same radii in Perseus \citep{schmidt02}, Hydra A, and Abell~1835 \citep{kirkpatrick11} and is consistent with some of the most metal-peaked clusters, including Abell~262 \citep{kirkpatrick11}. The fact that these three clusters at $\left<z\right> \sim 0.7$ have slightly ($\sim$1$\sigma$) over-enriched cores compared to those from \cite{baldi07} may indicate that the process responsible for mixing gas (e.g., AGN feedback, sloshing of the cool core in the cluster potential) in the central part of the cluster was not operating as effectively 6 Gyr ago as it is today.

While it has been well established that the bulk of the metals in the cluster ICM were formed early on \citep{werner13,ettori15,mcdonald16a,mantz17}, there are relatively few studies which have targeted the metal-enriched cores of cool core clusters, which are typically overabundant by $\Delta Z$ $\sim$ 0.3Z$_{\odot}$ \citep{degrandi01}. \cite{mantz17} found no evolution ($\beta_{1+z} = -0.14 \pm 0.17$) in the core ($0.0-0.1$R$_{500}$) metallicity of massive clusters spanning $0 < z < 1.2$, while \cite{mcdonald16a} find dZ/d$z$ $=$ $-0.04 \pm 0.1$ Z$_{\odot}$ for the cores ($0.0-0.15$R$_{500}$) of clusters spanning $0.1 < z < 1.7$. Considering only cool core clusters, \cite{mcdonald16a} find dZ/d$z$ $=$ $-0.21 \pm 0.11$ Z$_{\odot}$, which was suggestive of mild ($\sim$1/3 of metals in core created since $z\sim1$) evolution in the core metallicity. 

While this work is based on only 4 clusters, one of which is highly contaminated in the core due to the presence of a bright point source, it suggests that, for the most relaxed clusters, the central metal excess that we observe at $z\sim0$ was already in place 6 Gyr ago. We will return to this in a discussion below.

\subsection{Central Galaxy Properties}

\begin{figure*}[htb]
\centering
\includegraphics[width=0.99\textwidth]{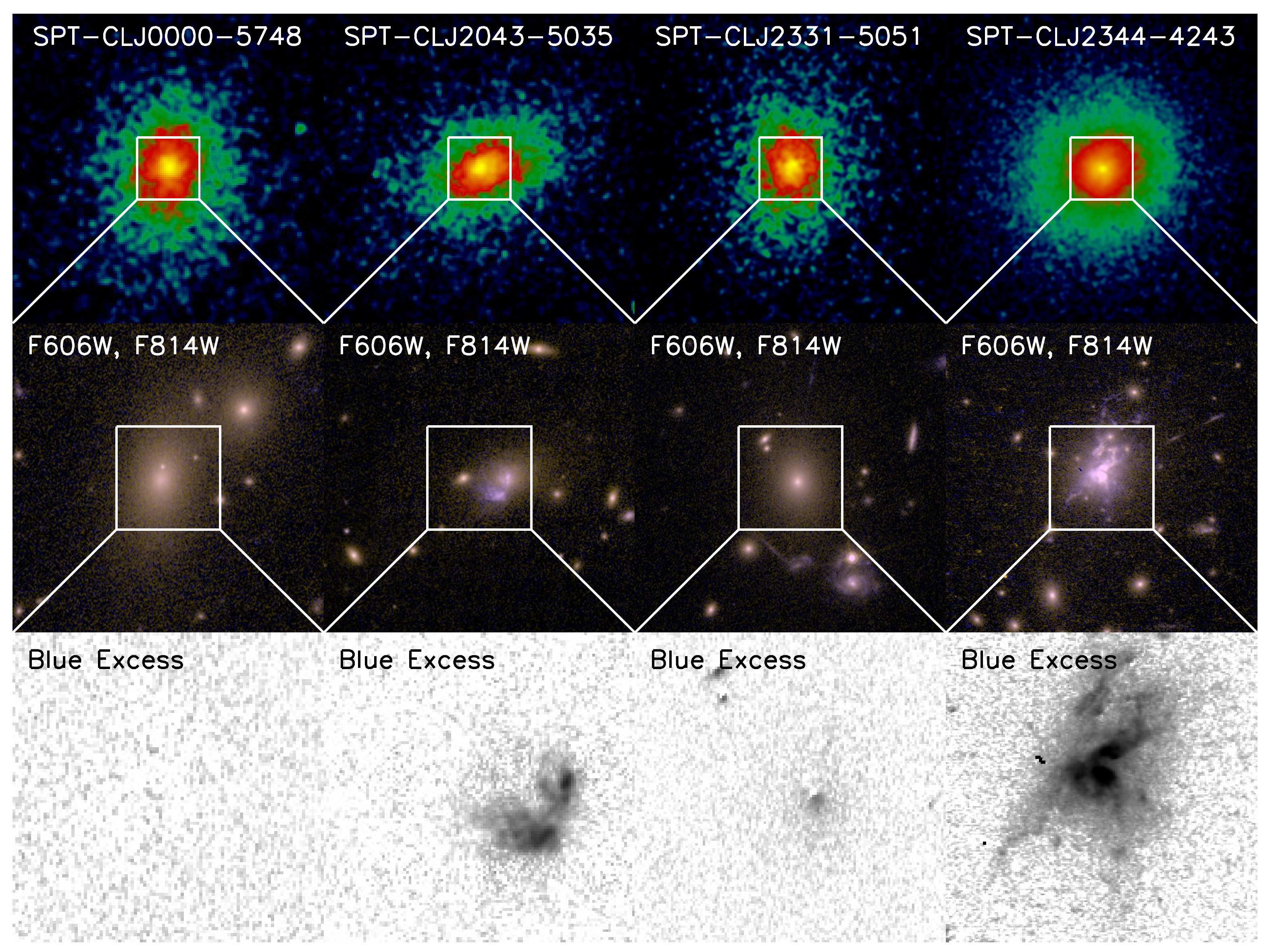}
\caption{\emph{Upper row:} X-ray image (0.7--4.0\,keV) of each cluster, showing the relatively relaxed, centrally-concentrated morphology. Images are 1\,Mpc on a side. \emph{Middle row:} Zoomed-in optical images of the cluster core, showing the central BCG in each cluster. These images are made by combining the F606W (blue) and F814W (green, red) images from HST in such a way as to make the old stellar populations appear red and the young populations blue. Both SPT-CLJ0000-5748 and SPT-CLJ2331-5051 appear relatively quiescent in these colors, while SPT-CLJ2043-5035 and SPT-CLJ2344-4243 clearly have significant younger stellar components. \emph{Lower row:} Zoomed in image showing the excess emission in the F606W band above the expectation for a passive stellar population, based on the F814W band image. These panels show an absence of emission for the quiescent (based on color) BCGs, and clumpy, extended emission for SPT-CLJ2043-5051 and SPT-CLJ2344-4243. This clumpy, blue emission, which is extended on scales of $\sim$15\,kpc (SPT-CLJ2043-5051) and $\gtrsim$50\,kpc (SPT-CLJ2344-4243), is further evidence for ongoing star formation.}
\label{fig:bcg_images}
\end{figure*}

Relaxed, cool core galaxy clusters at $z\sim0$ tend to have star-forming central galaxies \citep[e.g.,][]{mcnamara89,odea08,mcdonald18a}, which are well-aligned with the X-ray peak. This seems to be the case at higher redshift to the degree that it has been tested \citep[e.g.,][]{fogarty15}. Recent work \citep{webb15,mcdonald16b,bonaventura17} has shown that, at $z\gtrsim1$, clusters harbored central galaxies that were forming stars at rates of $\sim$100 M$_{\odot}$ yr$^{-1}$, compared to typical rates of 1--10 M$_{\odot}$ yr$^{-1}$ at $z\sim0$ \citep{odea08,mcdonald18a}. However, much of this high-$z$ star formation appears to be fueled by gas-rich mergers, and is predominantly found in the centers of disturbed, non-cool core clusters. This study represents an opportunity to test, in the most relaxed clusters at $z\sim0.7$, how much star formation can be attributed to cooling of the hot ICM.

In Figure \ref{fig:bcg_images}, we show X-ray and optical images of the cluster (upper panels, 1 Mpc on a side), the central core (middle panels, 200 kpc on a side), and the central galaxy (lower panels, 50 kpc on a side). In all four clusters, there is a massive, giant elliptical galaxy nearly coincident with the X-ray peak. We find physical, 2-dimensional offsets between the BCG and the X-ray peak of $\lesssim$3--10 kpc (see Table \ref{table:bcg}). These small offsets are consistent with what is found in typical low-$z$ cool core clusters \citep[$\sim$3--10\,kpc;][]{sanderson09}, and indicate a relatively small degree of dynamical activity. SPT-CLJ2331-5051 and SPT-CLJ2344-4243 appear to harbor the most dominant central galaxy, with no other massive galaxies within the inner $\sim$100\,kpc of the cluster center. SPT-CLJ0000-5748 has a close, massive companion that appears to be gas poor, and may be in the midst of merging with the central galaxy, though this could be a projection effect. Both SPT-CLJ0000-5748 and SPT-CLJ2331-5051 have overall very red colors, and show no sign of structure in the F606W (rest-frame blue) band, indicating relatively old stellar populations and little to no star formation.

Both SPT-CLJ2043-5035 and SPT-CLJ2344-4243 harbor central galaxies with excess clumpy blue emission, indicating significant ongoing star formation. While these systems are similar in terms of their stellar populations, they are very different in other ways. SPT-CLJ2043-5035 has the largest BCG offset from the X-ray peak, while SPT-CLJ2344-4243 has no measurable offset between the X-ray peak and the central galaxy. The BCG in SPT-CLJ2043-5035 is the least massive at M$_* = 4.1\times10^{11}$ M$_{\odot}$, while the BCG in SPT-CLJ2344-4243 is the most massive at M$_* = 14.5 \times 10^{11}$ M$_{\odot}$. The blue emission in SPT-CLJ2043-5035 is extended on scales of $\sim$15\,kpc and appears to be double-peaked, while SPT-CLJ2344-4243 is centrally concentrated in a single peak and extended on scales of $\gtrsim$50\,kpc.
In both cases, the young stars may be the result of a gas-rich merger, and/or cooling of the low-entropy gas at the center of the cluster. It is challenging to differentiate between these two scenarios with the available data -- the old stellar populations are not obviously disturbed in either system, but we also do not have a sufficiently red band with high enough angular resolution to definitely make such a statement. There is a potential donor galaxy to the east of the BCG in SPT-CLJ2043-5035, but it is quite small and symmetric -- to lose enough gas to fuel such a large amount of star formation, it would have to be fully disrupted. We will discuss these systems further in \S4.1.

\begin{figure}[htb]
\centering
\includegraphics[width=0.49\textwidth]{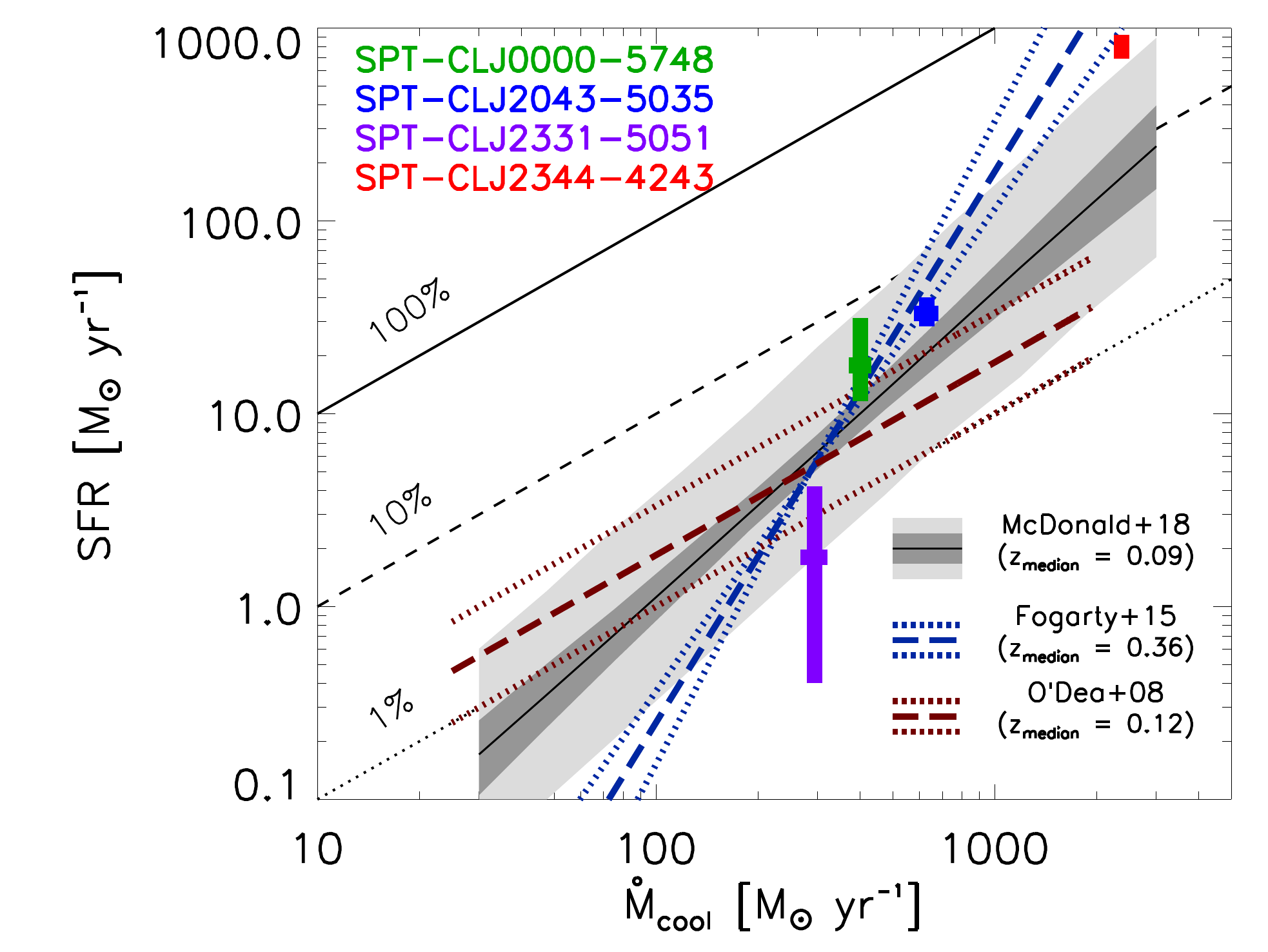}
\caption{BCG star formation rate compared to classical cooling rate (\.M$_{cool} \equiv M_g(r<r_{cool})/t_{cool}$) for the four clusters in our sample. These star formation rates are based on the [O\,\textsc{ii}] emission line, and are corrected for intrinsic extinction as measured from the CIGALE fit to the broadband SED. For comparison, we show the relations found in \citet[][dark red]{odea08} and \citet[][black/grey]{mcdonald18a} for clusters at $z\sim0.1$. For the latter, we show the uncertainty on the fit in dark grey, and the measured scatter in light grey. In blue, we show the fit to CLASH clusters from \cite{fogarty15}, which are at considerably higher redshift. Overall, the relaxed clusters in the SPT sample (including Phoenix, \citep{mcdonald12c}, which is the most relaxed SPT-selected cluster) tend to host BCGs with higher SFR per unit cooling ICM than clusters at $z\sim0$.
}
\label{fig:sfr_dmdt}
\end{figure}

In Figure \ref{fig:spectra}, we showed the optical spectra of the four BCGs in this sample. These BCGs represent four very different phases of galaxy evolution. SPT-CLJ2043-5035 and SPT-CLJ2344-4243 both have strong [O\,\textsc{ii}] emission and no evidence of a 4000\AA\ break, indicating that the light is dominated by young stellar populations. The specific star formation rates (sSFR) of these systems are 0.09 Gyr$^{-1}$ and 0.554 Gyr$^{-1}$, indicating that, while both are rapidly forming stars, this star formation is only contributing significantly to the growth of SPT-CLJ2344-4243, which will double its mass in $\sim$2\,Gyr (compared to $\sim$10\,Gyr for SPT-CLJ2043-5035).
SPT-CLJ0000-5748 also has notable [O\,\textsc{ii}] emission, but also exhibits a strong 4000\AA\ break and deep absorption lines, indicative of an old stellar population dominating the emission. The corresponding sSFR for this system is 0.014 Gyr$^{-1}$, indicating that star formation is contributing negligibly to the growth of the total stellar mass. This is consistent with Figure \ref{fig:bcg_images}, in which the light is dominated by the smooth, red stellar population. Finally, SPT-CLJ2331-5051 provides a fourth spectral type: no significant [O\,\textsc{ii}] emission, and an overall old stellar population (strong 4000\AA\ break, deep absorption lines). This BCG has sSFR of 0.002 Gyr$^{-1}$, indicating that it is evolving almost completely passively.

Despite living in nearly identical clusters (see Figures \ref{fig:xray_profs}, \ref{fig:selfsim}, \ref{fig:entropy}), these four BCGs span a factor of $\sim$500 in specific star formation rate. In Figure \ref{fig:sfr_dmdt}, we compare the cooling rate to the BCG star formation rate, following \cite{mcdonald18a}, where we are assuming that the star formation in the BCG is connected to the cooling of the hot ICM (we will address this assumption in \S4.1). While the host clusters span a factor of $<$10 in cooling rate, the BCGs span a factor of $\sim$500 in star formation rate, consistent with the considerable scatter in star formation rates at fixed cooling rate measured in \cite{mcdonald18a}. These four systems are consistent with both the trends found by \cite{fogarty15} for clusters at $z\sim0.4$ and \cite{mcdonald18a} for clusters at $z\sim0.1$. Both of these studies find a slope greater than unity, suggesting that the more massive, strongly cooling clusters are also cooling more efficiently.
For one out of the four clusters (SPT-CLJ2344-4243), the ratio of the star formation rate to the cooling rate is outside of the 1$\sigma$ scatter measured for a sample of $>$100 cool core clusters at $z\sim0$ \citep{mcdonald18a}, consistent with expectations.
If the star formation in all of these systems can be attributed to cooling, it suggests that, with the exception of the Phoenix cluster (SPT-CLJ2344-4243), cooling is suppressed as effectively at early times as it is today. We will discuss this further in \S4.1.
%

In summary, all four clusters host massive, central galaxies within $\lesssim$10\,kpc of the X-ray peak, consistent with low-$z$ observations of relaxed clusters. These central galaxies span a range of stellar populations, from completely passive (SPT-CLJ2331-5051) to rapidly star-forming (SPT-CLJ2043-5035), to starburst (SPT-CLJ2344-4243), despite living at the centers of very similar clusters.


\section{Discussion}

\subsection{What is the Origin of the Star Formation?}

%

Figures \ref{fig:spectra} and \ref{fig:bcg_images} demonstrate that, despite sharing similar properties on the cluster scale, the central BCGs in these four clusters could not be more different, spanning the full range from passive to starburst. In the cases of SPT-CLJ2043-5035 and SPT-CLJ2344-4243, the star formation rates imply the presence of a tremendous amount of cold gas. This cold gas is most likely the result of either cooling of the hot intracluster medium, so-called ``residual cooling flows'' \citep[see e.g.,][]{mcdonald18a}, or stripping of gas-rich galaxies as they pass through the dense cluster core. In this section we will attempt to differentiate between these two scenarios given the available data.

The morphology of the blue excess in SPT-CLJ2043-5035 (Figure \ref{fig:bcg_images}) is more similar to that of a gas-rich merger than of a typical cool core cluster. The extended blue emission points towards a smaller red galaxy to the east of the BCG, consistent with a scenario in which a satellite galaxy was stripped while passing close to the central BCG. However, such a double-peaked morphology in the blue excess could also be a result of a sloshing cool core. There are several low-$z$ systems where the cool core has been ``dislodged'' from the BCG by a minor interaction, leading to the condensation of low-entropy gas away from the direct influence of the AGN \citep{hamer12}. In this case, a minor merger would be responsible for setting the core in motion, but the cool gas would originate in the hot phase, not in a donor galaxy. We will investigate further the relationship between the dynamical state of these clusters and the properties of the BCG and their AGN in a companion paper.

\begin{figure}[t]
\centering
\includegraphics[width=0.49\textwidth]{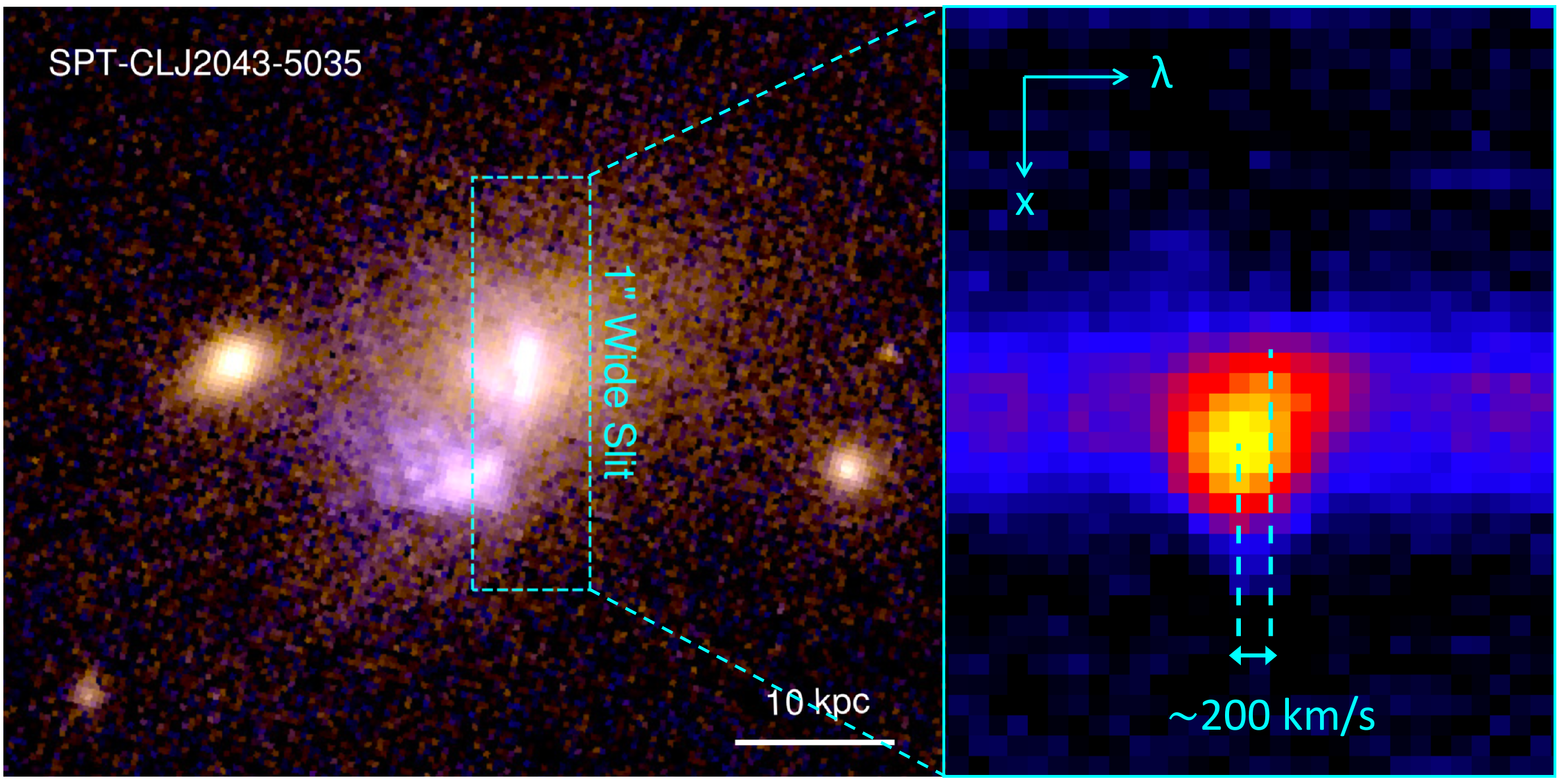}
\caption{\emph{Left:} Pseudo-color image, combining the F606W and F814W filters on HST, of the central galaxy in SPT-CLJ2043-5035. The cyan rectangle shows the position of the slit that was used to obtain the optical spectrum (Figure \ref{fig:spectra}). \emph{Right:} Two-dimensional spectrum, extracted along the slit shown in the left panel, and centered on the [O\,\textsc{ii}]$\lambda\lambda$3726,3729 doublet. This spectrum shows that the emission peak shifts by $\sim$200\,km/s over the extent of the extended line emission.
}
\label{fig:clj2043}
\end{figure}

Conveniently, the N-S oriented long-slit that was placed on the BCG captures much of the extended blue emission, as we show in Figure \ref{fig:clj2043}. The two-dimensional spectrum, shown in the right panel of Figure \ref{fig:clj2043} shows extended [O\,\textsc{ii}] emission, with a velocity gradient of $\sim$200\,km/s across the extended emission. For comparison, the velocity spread observed for stripped galaxies in dense environments from the GASP (Gas Stripping Phenomena) survey is significantly higher, with measured values of $\Delta v \sim 500$ km/s \citep[J0201, J0206;][]{poggianti17a, poggianti17b}, consistent with the velocity dispersions measured in the host cluster. Given that the clusters considered here are more massive than those from the GASP survey, we would expect even higher velocity widths across the length of the stripped gas. However, if the stripping is happening in the plane of the sky, the velocity spread could be significantly diminished, to (or below) the levels observed here. For comparison, the velocities spanned by cooling, multiphase gas in the cores of nearby clusters are $\sigma_v \sim 100-200$ km/s \citep[e.g.,][]{gaspari18}, fully consistent with what we observe in SPT-CLJ2043-5035.

%

Beyond the morphology and dynamics, we can consider the amount of star formation, and whether it could realistically be fueled by stripping. For SPT-CLJ2043-5035, the nearest galaxy (east of the BCG in Figure \ref{fig:clj2043}) is the most likely donor, and has a stellar mass $\sim$5$\times$ smaller than the BCG, based on its $i$-band brightness. Combining the amount of star formation with the stellar mass of the donor galaxy yields a specific star formation rate of 0.4 Gyr$^{-1}$. For comparison, this is only slightly higher than typical stripped dwarf galaxies in the GASP survey \citep[$sSFR  \sim 0.2-0.3$ Gyr$^{-1}$;][]{vulcani17,george18}, and much less than we see in starburst galaxies such as M82 and NGC1569 \citep[sSFR $\sim$ 1.0 Gyr$^{-1}$;][]{jarrett17}. For SPT-CLJ2344-4243, the implied sSFR in the nearest potential donor galaxy is $\sim$10 Gyr$^{-1}$, which is an order of magnitude higher than the most vigorous starbursts that we observe \citep[e.g.,][]{jarrett17}. As discussed in \cite{mcdonald12c}, this is strong evidence against the fueling of star formation by the stripping of infalling gas in this system, as it would require several ($\sim$10) gas-rich galaxies all simultaneously donating their gas.

Finally, there is evidence provided by the cooling properties of the ICM. The rank order of the cooling rate and the star formation rates are identical -- i.e., the strongest cool core harbors the most star-forming BCG, the second strongest cool core has the second most star-forming BCG, etc. If the star formation is unrelated to cooling, this would happen by chance $<$5\% of the time. Both SPT-CLJ2043-5035 and SPT-CLJ2344-4243, which harbor BCGs forming stars on large physical scales ($>$10\,kpc), have entropy profiles that fall below the gravity-only prediction \citep{voit05} between 30--100\,kpc, while the other two clusters (which show no evidence of extended star formation) lie above it at all radii. The facts that both the entropy profiles and BCG stellar populations divide the four clusters into the same two groups suggest that the star formation is fed by the cooling ICM, though this is not conclusive.

Indeed, none of these arguments alone provide conclusive evidence for a cooling, rather than stripping, origin for the star-forming gas. While it seems most probable that SPT-CLJ2344-4243 is cooling-fed, based solely on the overwhelming amount of available cold gas that would have to come from several donors, the picture is not so clear for SPT-CLJ2043-5035. The morphology appears to favor a stripping origin, while the thermodynamic profiles seem to favor cooling. The dynamics and total amount of star formation do not strongly favor either interpretation. With deeper X-ray data we could look for a spatial correlation between the low-entropy gas and the star formation, while deeper, ideally spatially resolved (i.e., IFU), optical spectroscopy could allow us to investigate the kinematics and metallicity of the young stars, and whether these are more similar to the cooling ICM or the nearest donor galaxy.

\subsection{Metal Enrichment in Cluster Cores}
One of the leading explanations for the centrally-peaked metallicity profile is that the BCG has enriched the ICM in the immediate vicinity via type Ia supernovae over several Gyr \citep[e.g.,][]{degrandi04}. 
In such a scenario, we may expect a significant change in the magnitude of the central metallicity excess between $z\sim0.7$ and $z\sim0$, which represents 6\,Gyr, or nearly half of the age of the Universe. We can calculate the expected type Ia supernova rates from the central galaxy between $z=2$ (roughly the cluster formation time) and $z=0.7$, and then $z=0.7$ and $z=0$, to estimate roughly what fraction of the central metallicity excess was formed at late times, if this is indeed the enrichment mechanism. We assume supernova rates (SNR) from \cite{perrett12}, which account for both prompt SN shortly after the formation of massive stars (scaling with SFR), and delayed SN which occur much later (scaling with stellar mass):

\begin{equation}
SNR_{Ia}(z) = 1.9 \times 10^{-14} M_*(z) + 3.3 \times 10^{-4} SFR(z) ~,
\end{equation}

\noindent{}where we take the BCG stellar mass as a function of redshift from \cite{delucia06}, with M$_* = 5\times10^{11}$ at $z=0$, and the BCG star formation rate as a function of redshift from \cite{bonaventura17}. We find that 24\% of type Ia supernovae in BCGs should have exploded at $z<0.7$, with the bulk of the enrichment in the core happening at $z>0.7$. We note that this does not account for core collapse SNe, which likely dominated at early times when BCGs were exceptionally star-forming \citep[see][]{mcdonald16b,bonaventura17}. Thus, we expect this to represent an upper limit on the fraction of metals produced in cluster cores at $z<0.7$ compared to $z>0.7$.
Based on this simple Ia-only model, we expect $\sim$50\% of the core enrichment to happen between $z=2$ and $z=1.5$, and so it is unsurprising that we, and previous studies \citep[e.g.,][]{mantz17}, do not see a strong evolution in the central metallicity excess.

Perhaps more important in dictating the shape of the metallicity profile at late times ($z<1$) is the central AGN. \cite{kirkpatrick11} showed that radio-loud AGN in the centers of clusters can push metals from the central core to large radii ($\gtrsim$100\,kpc). 
Figure \ref{fig:metallicity} shows that all three clusters for which we can constrain the central metallicity have a metallicity peak in the inner $\sim$10\,kpc that is over-enriched at the 1$\sigma$ level when compared to the average profile for low-redshift cool core clusters from \cite{baldi07}. If pushed to radii of 50--100\,kpc (a volume thousands of times larger) these metals would quickly be diluted, and the metallicity profile would be indistinguishable from the \cite{baldi07} profile. Thus, the presence, or lack, of a sharply peaked metallicity profile may be telling us more about the amount of time elapsed since the last major outburst of AGN feedback than it is about the enrichment history of the cluster core. Given how centrally concentrated the metallicity profiles are in these three systems, it is likely that neither has experienced a major outburst in a few hundred million years, which corresponds to the free fall time at a radius of $\sim$100\,kpc. This is corroborated by the fact that the observed bubbles in these systems are at relatively small radii, indicating ongoing, rather than past, feedback \citep{hlavacek15}. We will investigate this scenario further in a companion paper, focusing specifically on the feedback and dynamical properties of these four clusters.

\section{Summary}

We present new data from the \emph{Chandra X-ray Observatory} and the \emph{Hubble Space Telescope}, targeting the four most relaxed clusters in the initial South Pole Telescope 2500 deg$^2$ survey. These represent some of the deepest data currently available for clusters at $z>0.5$. In this work, we focus on the cooling properties of the intracluster medium, along with the stellar populations of the central brightest cluster galaxy. We find:

\begin{itemize}

\item The thermodynamic profiles of all four clusters are very similar to one another and to clusters at $z\sim0$. This includes the shape of the temperature profile, which is well described by the universal model \citep{vikhlinin06a}, and the entropy profile, which is well described by the ensemble profiles for clusters at $z\sim0$ \citep[e.g.,][]{walker12,panagoulia14,babyk18}. We find no evidence for deviations from self similar evolution in the temperature profiles, implying that the process responsible for preventing runaway cooling over the past $\gtrsim$6\,Gyr is preserving self similarity.
We compare the measured thermodynamic profiles to those published in \cite{mcdonald13b} -- based on data a factor of $\sim$5 shallower -- and find good agreement, suggesting that the assumptions made when interpreting low S/N data (e.g., shape of temperature profile, constant metallicity, fixed redshift) are valid for relaxed clusters.

\item Despite representing 6 Gyr in evolution between our sample and well-studied low-$z$ clusters, we see no evidence for a change in the cooling properties of the core, with central temperature drops of 0.15--0.4 \citep[compared to typical values of 0.1--0.4 for cool core clusters at $z\sim0$;][]{vikhlinin06a}, central ($r\sim5$\,kpc) entropies of 11--16 keV cm$^2$ \citep[compared to typical values of $\sim$15 keV cm$^2$ for cool core clusters at $z\sim0$;][]{panagoulia14}, and central cooling times of 0.18--0.32\,Gyr \citep[compared to typical values of 0.3--0.8 for cool core clusters at $z\sim0$][]{hogan17}. This implies a tight balance between heating and cooling over the past $\geq$6\,Gyr.

\item We find that the metallicity of the ICM in both the central region ($r<0.1R_{500}$) and core-excised region ($0.1-0.5R_{500}$) agree well with what is found at $z\sim0$. This adds further evidence for early enrichment of the ICM. Interestingly, we find mild (1$\sigma$) evidence for \emph{over}-enriched cores at $z\sim0.7$ compared to $z\sim0$. We calculate that the bulk ($>$76\%) of metallicity excess observed at the centers of clusters today came from supernovae at $z>0.7$, confirming that we should not expect to see a strong evolution in the central metal excess over the past 6\,Gyr. We propose that, instead, the variations in central metallicity are telling us more about the timescales of strong AGN feedback (which can redistribute metals). This would imply that the three systems for which we constrain the inner metallicity here have not experienced a major AGN outburst, capable of pushing metals outside of $\sim$100\,kpc, in the last few hundred million years. 

\item Despite sharing remarkably similar cooling properties (e.g., central cooling time, classical cooling rate), the central galaxies in these four clusters exhibit markedly different stellar populations, ranging from completely passive (SPT-CLJ2331-5051: no emission lines, strong 4000\AA\ break), to weakly star-forming (SPT-CLJ0000-5748: weak emission lines, strong 4000\AA\ break), to strongly star-forming (SPT-CLJ2043-5035: strong emission lines, weak 4000\AA\ break), to starburst (SPT-CLJ2344-4244: young stellar populations dominate emission). 
If all of this star formation is due to cooling of the hot ICM (which may not be the case for SPT-CLJ2043-5035), it implies that the relationship  between the cooling rate and star formation rate at early times is similar to that observed for nearby clusters, with considerable scatter in star formation at fixed cooling rate and a steeper-than-unity slope in the star formation rate as a function of cooling rate.

\end{itemize}

\noindent{}This analysis provides a reference point for our past and future analyses of distant clusters. Observations of such systems are, by necessity, typically shallow, requiring leaps of faith in interpreting unresolved ground-based data or low-count X-ray data. 
With these deep, high angular resolution data, we can anchor these analyses at the halfway point of cluster evolution, providing confidence when future observations extend these measurements even further into the past.


\section*{Acknowledgements} 
%
Support for this work was provided by NASA through Chandra Award Numbers GO6-17112A and GO6-17112B issued by the Chandra X-ray Observatory Center, which is operated by the Smithsonian Astrophysical Observatory for and on behalf of the National Aeronautics Space Administration under contract NAS8-03060.
Additional support was provided by NASA through grants from the Space Telescope Science Institute (HST-GO-13578, HST-GO-14352), which is operated by the Association of Universities for Research in Astronomy,
Incorporated, under NASA contract NAS5-26555. 
The South Pole Telescope is supported by the National Science Foundation through grant PLR-1248097. Partial support is also provided by the NSF Physics Frontier Center grant PHY-1125897 to the Kavli Institute of Cosmological Physics at the University of Chicago, the Kavli Foundation and the Gordon and Betty Moore Foundation grant GBMF 947.
SWA and ABM acknowledge additional support from the U.S. Department of Energy under contract
number DE-AC02-76SF00515.
JHL is supported by NSERC through the discovery grant and Canada Research Chair programs.
WF acknowledges support from NASA contract NAS-08060 and the Smithsonian Institution.
The Munich group acknowledges the support by the DFG Cluster of Excellence ``Origin and Structure of the Universe'', the Ludwig-Maximilians-Universit\"at (LMU-Munich), and the Transregio  program TR33 ``The Dark Universe''.
TS acknowledges support from the German Federal Ministry of Economics and Technology (BMWi) provided through DLR under projects 50 OR 1210, 50 OR 1407 and 50 OR 1610.
The Melbourne group acknowledges support from the Australian Research Council's Discovery Projects funding scheme (DP150103208).
BB is supported by the Fermi Research Alliance, LLC under Contract No. De-AC02-07CH11359 with the U.S. Department of Energy.


\end{document}